\documentclass[aps,showpacs,nofootinbib,superscriptaddress,twocolumn]{revtex4}

\usepackage{graphicx}
\usepackage{bm}
\usepackage{amsmath}
\usepackage{amssymb}
\usepackage{hyperref}

\newcommand{\nslash}{\kern 0.2 em n\kern -0.50em /}
\newcommand{\kslash}{\kern 0.2 em k\kern -0.45em /}
\newcommand{\pslash}{\kern 0.2 em p\kern -0.50em /}
\newcommand{\Sslash}{\kern 0.2 em S\kern -0.50em /}
\newcommand{\Pslash}{\kern 0.2 em P\kern -0.50em /}
\newcommand{\Dslash}{\kern 0.2 em D\kern -0.65em /\kern 0.15em}
\newcommand{\slim}{\mskip 1.5mu}

\begin{document}

\title{Azimuthal asymmetries in single polarized proton-proton Drell-Yan processes}


\newcommand*{\SEU}{Department of Physics, Southeast University, Nanjing
211189, China}\affiliation{\SEU}
\newcommand*{\UTFSM}{Departamento de F\'\i sica, Universidad T\'ecnica
Federico Santa Mar\'\i a, and Centro Cient\'\i fico-Tecnol\'ogico de
Valpara\'\i so Casilla 110-V, Valpara\'\i so,
Chile}\affiliation{\UTFSM}
\newcommand*{\PKU}{School of Physics and State Key Laboratory of Nuclear Physics and
Technology, \\Peking University, Beijing 100871,
China}\affiliation{\PKU}
\newcommand*{\CHEP}{Center for High Energy
Physics, Peking University, Beijing 100871,
China}\affiliation{\CHEP}

\author{Zhun Lu}\affiliation{\SEU}\affiliation{\UTFSM}
\author{Bo-Qiang Ma}\email{mabq@pku.edu.cn}\affiliation{\PKU}\affiliation{\CHEP}
\author{Jiacai Zhu}\affiliation{\PKU}

\begin{abstract}
We study the azimuthal asymmetries in proton-proton Drell-Yan
processes with one incident proton being transversely or
longitudinally polarized. We consider particularly the asymmetries
contributed by the leading-twist chiral-odd quark distributions. We
analyze the asymmetries with $\sin(2\phi+\phi_S)$ and
$\sin(2\phi-\phi_S)$ modulations in transverse single polarized
$p^\uparrow p$ Drell-Yan and $\sin2\phi$ asymmetries in longitudinal
single polarized $p^\rightarrow p$ Drell-Yan at the Relativistic Heavy Ion
Collider, the Japan Proton Accelerator Research Complex, E906 (Fermi
Lab), and the Nuclotron-based Ion Collider Facility (Joint Institute for
Nuclear Research). We show that the measurements of
the asymmetries in those facilities can provide valuable information
of the chiral-odd structure of the nucleon both in the valence and
sea regions.
\end{abstract}


\maketitle

\section{Introduction}

The single spin asymmetry (SSA) appearing in various high-energy
scattering
processes~\cite{bdr,D'Alesio:2007jt,Barone:2010ef,Boer:2011fh} is
among the most challenging issues of QCD spin physics. Large SSAs
were observed experimentally in the process $p \, p^{\uparrow}
\rightarrow \pi \, X$ \cite{Adams} two decades ago. Standard
perturbative QCD based on collinear factorization to leading power
of $1/Q$ cannot explain these asymmetries~\cite{Kane}. Many
theoretical studies~\cite{boros,sivers,anselmino95} have been
proposed to explain the origin of such asymmetries. One standard
approach is to assume the existence of parton distribution and/or
fragmentation depending on intrinsic transverse momentum, by going
beyond the collinear picture. In this transverse momentum dependent
(TMD) framework, novel structures of the nucleon emerge. For
instance, due to the correlation of nucleon transverse spin $\bm{S}$
and quark transverse momentum $\bm{k}_T$, there can be an asymmetric
distribution of unpolarized quarks in a transversely polarized
proton~\cite{Bacchetta:2004jz}:
 \begin{align}
&f_{q/p^\uparrow}(x, {k}_T) - f_{q/p^\uparrow}(x, -{k}_T)\nonumber\\
=& \Delta^N f_{q/p^\uparrow}(x, k_T^2) \, \frac{(\hat{\bm P}
  \times {\bm k}_T) \cdot {\bm S}}{|{\bm k}_T|} \nonumber\\
=& -2
 \, f_{1T}^{\perp q}(x, k_T^2)\frac{(\hat{\bm P}
  \times \, { \bm k}_T) \cdot {\bm S}}{M} .
\end{align}
Here $f_{1T}^{\perp q}$ or $\Delta^N f_{q/p^\uparrow}$ is referred
to as the Sivers function~\cite{sivers,anselmino95}, and has been
applied to explain the SSAs  observed in the process $p \,
p^{\uparrow} \rightarrow \pi \, X$. For a while the Sivers function
was thought to be forbidden by the time-reversal invariance property
of QCD~\cite{collins93}. However, model calculations~\cite{bhs02} by
Brodsky, Hwang and Schmidt show that the Sivers effect can be
allowed in the semi-inclusive deeply inelastic scattering (SIDIS)
and Dell-Yan process at leading-twist level, due to the
final/initial state interaction (FSI/ISI) between the struck quark
and the target remnant. It was then realized that FSI/ISI can be
accumulated into the Wilson lines (gauge-links) that are the key
ingredients for a full gauge-invariant definition~\cite{jy02,bmp03}
of TMD distribution functions. This also leads to the prediction on
the sign reversal of the Sivers functions in SIDIS and
Drell-Yan~\cite{collins02}.
For hadron productions in hadron-hadron collision (i.e., $H_A+ H_B \rightarrow h_1 + h_2 +X$),
the situation is more involved, as there are colored objects in both the initial
state and the final state.
The multiple FSI/ISI will generate process-dependent TMD distributions~\cite{cq07,collins07,vy07,bmnpb08}
which are different from those in SIDIS or Drell-Yan process. This is also viewed as
the breakdown of the generalized TMD factorization in inclusive hadro-production
of hadrons~\cite{Rogers:2010dm}.

Allowing naive-$T$-odd parton distributions encourages a lot of theoretical and experimental studies.
Substantial SSAs contributed by the Sivers effect in SIDIS processes
~\cite{smc,Airapetian:2004tw,compass,hermes05,compass06,2009ti,Alekseev:2010rw,Qian:2011py},
with one colliding nucleon transversely polarized, have been
measured by several experiments during recent years.
The asymmetries are identified by the angular
dependence $\sin(\phi_h-\phi_S)$, where $\phi_h$ and $\phi_S$ denote,
respectively, the azimuthal angles of the produced hadron and of the
nucleon spin polarization, with respect to the lepton scattering
plane. The data on the Sivers SSAs have been utilized by
different groups~\cite{anselmino05b,efr05,cegmms,vy05,Anselmino:2008sga}
to extract the Sivers functions of the proton, on the basis of the TMD
factorization~\cite{Ji:2004xq,col04}. Those sets of parametrizations
of the Sivers functions were applied to
predict the Sivers SSA in various processes, such as the SIDIS at Jefferson Lab (JLab), and the Drell-Yan
processes at the COmmon Muon Proton Apparatus for Structure and Spectroscopy, the Relativistic Heavy Ion Collider (RHIC), and  the Polarized
Antiproton eXperiment (PAX). Many planned measurements of SSAs in single polarized Drell-Yan processes
at the established or planned hadron accelerators/colliders have been proposed. One of the main goals of these experiments is to test the sign change of the Sivers functions in SIDIS and Drell-Yan process~\cite{efr05,Anselmino2009}, as a crucial prediction of QCD dynamics.
It is also worthwhile to mention that a sign mismatch for the $k_T$-moments of Sivers functions
has been found when the authors of Ref.~\cite{Kang:2011hk} compared the functions
extracted from SIDIS data and those extracted
from $p^\uparrow p \rightarrow \pi X$ data.

The planned polarized Drell-Yan processes at (future) available facilities also provide great opportunities to investigate various spin and transverse momentum dependent (TMD) distributions. Besides the Sivers effect, there are some other effects that may contribute to the azimuthal spin asymmetries at leading twist thereby could be measured in single polarized Drell-Yan processes. It is interesting to point out that all these leading-twist effects (except the Sivers effect) involve the chiral-odd parton distribution functions. For example, the following combinations
\begin{eqnarray}
h_1\otimes h_1^\perp,~~~h_{1T}^\perp \otimes h_1^\perp,~~~h_{1L}^\perp \otimes h_1^\perp, \label{chiodd}
\end{eqnarray}
will lead to SSAs with $\sin(2\phi-\phi_S)$, $\sin(2\phi+\phi_S)$
and $\sin2\phi$ angular dependences, respectively. Here $\phi$ and
$\phi_S$ are the azimuthal angles of the dilepton pair and proton
transverse spin with respect to the hadron plane, and we use the
convention for the angle definition introduced in
Ref.~\cite{Arnold2009}. These types of the asymmetries arise from
the coupling of two different chiral-odd parton distributions. The
coupling $h_1\otimes h_1^\perp$ was first introduced and analyzed in
Ref.~\cite{Boer1999} as an alternative mechanism for SSA and a
method of accessing the transversity distribution functions
$h_1$~\cite{Ralston:1979ys,Jaffe:1991kp}. The key ingredient for
these SSAs is the Boer-Mulders function $h_1^\perp$~\cite{bm}, which
is also a naive-$T$-odd TMD distribution function and provides the
necessary phase required for SSA. In this paper, we will present a
phenomenological analysis of these SSAs in the proton-proton
Drell-Yan process contributed by various leading-twist chiral-odd
distribution functions. We consider proton-proton induced polarized
Drell-Yan process, since there are several hadron
accelerators/colliders, such as RHIC, the Japan Proton Accelerator
Research Complex (J-PARC), E906 at Fermi Lab, and the Nuclotron-based
Ion Collider fAcility (NICA) at the Joint Institute for Nuclear
Research (JINR), that can perform these experiments. Therefore the
asymmetries at different energies and kinematical regions can be
analyzed and compared, which is important for obtaining the
information of various chiral-odd distributions functions from
experiments.

The remaining content of the paper is organized as follows. In
Sec. II, we briefly review the systematics of leading-twist
chiral-odd TMD quark distributions, then give the expressions of the
corresponding azimuthal angle weighted asymmetries
$A_{TU}^{\sin(2\phi-\phi_S)}$, $A_{TU}^{\sin(2\phi+\phi_S)}$ and
$A_{LU}^{\sin2\phi}$ in the framework of TMD factorization. We
consider both the single longitudinally and transversely polarized
Drell-Yan processes. In Sec. III, we present the phenomenological
predictions for single transverse spin asymmetry in $p^\uparrow p$
Drell-Yan process, and single longitudinal spin asymmetry in
$p^\rightarrow p$ Drell-Yan process at RHIC, J-PARC, E906 and NICA.
We conclude our paper in Sec. IV.

\section{Systematics of leading-twist chiral-odd distributions and their roles in SSAs}

At leading twist, according to the hermiticity properties of the fields and parity invariance, one may decompose the
TMD quark-quark correlation matrix of the
nucleon as follows~\cite{Mulders:1995dh,bm,Bacchetta:2006tn,Goeke2005}
\begin{widetext}
\begin{align}
\Phi(x,\bm{k}_T) &= \frac{1}{2}\, \biggl\{
f_1 \nslash_+
- {f_{1T}^\perp}\, \frac{\epsilon_T^{\rho \sigma} k_{T\rho}^{}\slim
  S_{T\sigma}^{}}{M} \, \nslash_+
+ \left(S_L\,g_{1L} - \frac{k_T \cdot S_T}{M}\,g_{1T}\right) \gamma_5\nslash_+
\nonumber \\[0.2em] & \quad \qquad
+h_{1T}\,\frac{\bigl[\Sslash_T, \nslash_+ \bigr]\gamma_5}{2}
+ \left(S_L\,h_{1L}^\perp - \frac{k_T \cdot S_T}{M}\,h_{1T}^\perp\right) \,\frac{\bigl[\kslash_T, \nslash_+ \bigr]\gamma_5}{2 M}
+i \, {h_1^\perp} \frac{ \bigl[\kslash_T, \nslash_+ \bigr]}{2M}
\biggr\}.
\label{eq:cor}
\end{align}
\end{widetext}
Here $n_+ = (0,1,\bm{0}_T)$ is a lightlike vector expressed in the
light-cone coordinates, in which an arbitrary four-vector $a$ is
written as $\{a^-,a^+, \boldsymbol{a}_T\}$, with $a^{\pm}=(a^0 \pm
a^3)/\sqrt{2}$ and $\boldsymbol{a}_T =(a^1,a^2)$.
The eight functions on the right-hand side of Eq.~(\ref{eq:cor}) not only depend on longitudinal
momentum fraction $x$, but also on the intrinsic transverse momentum of the quark
$\boldsymbol{k}_T$. Therefore they are named as transverse momentum dependent (TMD) distributions, or
alternatively the three-dimensional parton distribution
functions (3dPDFs) in momentum space. As the extensions of the usual
Feynman distribution functions, 3dPDFs enter the description of various
semi-inclusive reactions and encode a wealth of new information on the nucleon
structures that cannot be
described merely by the leading-twist collinear picture.

\begin{figure}[t]
\begin{center}
\includegraphics[width=0.40\columnwidth]{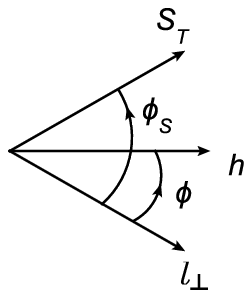}~~~~~~~~~
\includegraphics[width=0.40\columnwidth]{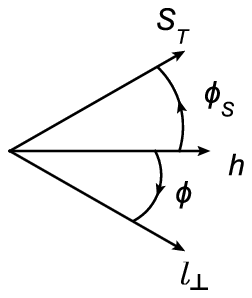}
\caption{\small Left panel: definition of azimuthal angles in Refs. ~\cite{Boer1999} and ~\cite{Bacchetta:2010si};
right panel: definition of azimuthal angles in Ref. ~\cite{Arnold2009}. Using the replacements $\phi\rightarrow -\phi$ and $\phi_S\rightarrow \phi_S-\phi$, the definition in the left panel is transformed to that in the right panel. }\label{angledef}
\end{center}
\end{figure}

Each of these eight 3dPDFs represents a special parton structure of
the nucleon. Five of them, the Sivers function $f_{1T}^\perp$, the
Boer-Mulders function $h_1^\perp$, the pretzelosity $h_{1T}^\perp$,
the transversal helicity $g_{1T}$, and the longitudinal transversity
$h_{1L}^\perp$, vanish upon integrating $\Phi(x,\bm{k}_T)$ over
$\boldsymbol{k}_T$. Particularly, two 3dPDFs, the Sivers function
and the Boer-Mulders function are naive-$T$-odd distributions and
account for the SSAs in various processes. Among the eight 3dPDFs,
$h_{1T}$, $h_1^\perp$, $h_{1T}^\perp$, and $h_{1L}^\perp$ are
chirally odd, that is, they describe densities of the probed quarks
with helicity flipped. Except $h_1^\perp$, other three chiral-odd
distribution are $T$-even. The relation between $h_{1T}$ given
in Eq.~(\ref{eq:cor}) and $h_1$ is
\begin{eqnarray}
h_1(x,k_T^2) = h_{1T}(x,k_T^2) + {\bm{k}_T^2 \over 2M^2} h_{1T}^\perp(x,k_T^2).
\end{eqnarray}
Since $h_1$ naturally appears in the expression of related azimuthal asymmetries, our discussion on the transversity
in the rest of our paper is based on $h_1$ rather than $h_{1T}$. The distributions $h_1$ and $h_1^\perp$ describe the densities of transversely polarized quarks inside a transversely polarized proton and an unpolarized proton, respectively.
The distributions
$h_{1T}^\perp$ and $h_{1L}^\perp$
 arise from double
spin correlations in the parton distribution functions (PDFs), representing the densities of transversely polarized quarks in a transversely
(but in a different direction) polarized proton and a longitudinally
polarized proton, respectively.

Because of the chiral-odd nature of $h_1$, $h_1^\perp$, $h_{1T}^\perp$
and $h_{1L}^\perp$, in high-energy processes they have to combine
together with another chiral-odd object, i.e., with the Collins
fragmentation function in SIDIS, or with another chiral-odd
distribution function in Drell-Yan, to manifest their effects. This
makes them rather difficult to be probed experimentally. As a
result, they are less known than the chiral-even distribution
functions. Anyway, there are some efforts to extract transversity
from SIDIS data~\cite{Anselmino:2007fs,Anselmino:2008jk}, and
Boer-Mulders function from SIDIS and Drell-Yan
data~\cite{Zhang:2008nu,Lu2009,Barone:2009hw,Barone:2010gk}. For
$h_{1T}^\perp$ and $h_{1L}^\perp$, there are extensive model
calculations~\cite{Pasquini:2008ax,Bacchetta:2008af,Avakian:2008dz,Shejun2009,
Efremov:2009ze,Boffi:2009sh,Avakian:2010br,Zhu2011,Zhu:2011ir}
and some proposals to measure them in SIDIS and $p\bar{p}$ Drell-Yan
processes.

All the leading-twist chiral-odd parton distributions can be probed in single polarized proton-proton Drell-Yan processes:
\begin{eqnarray}
p^{\uparrow/\rightarrow}(P_1) + p (P_2) \to \gamma^*(q)+X \to \ell(l) + \bar{\ell}(l^\prime) +X. \label{ppdy}
\end{eqnarray}
Here we assume that one proton (with momentum $P_1$) is polarized,
and $\uparrow$ or  $\rightarrow$ denotes its transverse polarization
or longitudinal polarization. In leading order, the dilepton pair is
produced from the annihilation of the quark and antiquark from each
proton. We denote the momenta of the annihilating partons from
polarized proton and unpolarized proton as $k_1$ and $k_2$,
respectively.  Then we can define the kinematical variables as
\begin{eqnarray}
q&=&l+l^\prime =(q^0,\bm{q}_T, q^3),~~Q^2 =q^2,~~ \nonumber\\
x_1&=&{Q^2 \over 2P_1\cdot q}\approx {k_1^+\over P_1^+},~~ x_1={Q^2 \over 2P_2\cdot q}\approx {k_2^-\over P_2^-},\nonumber\\
 y&=&{1\over 2}\ln\left({x_1\over x_2}\right).
\end{eqnarray}

\begin{figure*}[t]
\begin{center}
\includegraphics[width=0.8\textwidth]{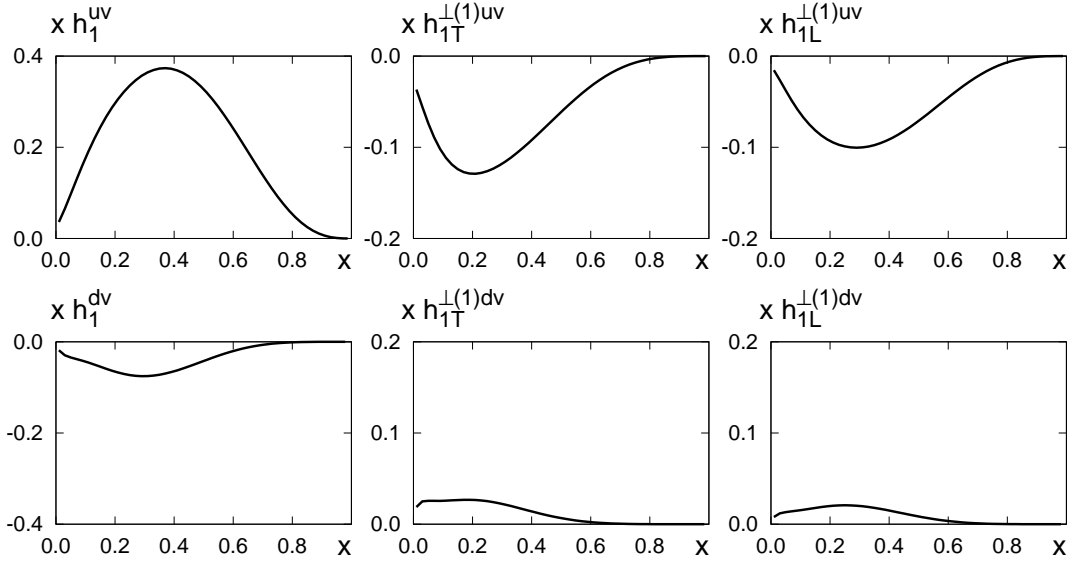}
\caption{The light-cone diquark model results of $x h_1(x)$ (left panels), $x h_{1T}^{\perp(1)}(x)$ (central panels), and $x h_{1T}^{\perp(1)}(x)$ (right panels) of valence $u$ and $d$ quarks at $Q^2 = 1~\mathrm{GeV}^2$, respectively.}\label{fig:teventmd}
\end{center}
\end{figure*}

In the Drell-Yan process,  if the transverse
momentum of the dilepton $\bm{q}_T$ is measured, we can apply the
TMD factorization~\cite{Collins1985,Ji:2004xq,col04} which is valid in the region $q_T^2\ll
Q^2$ to write down the differential cross section of processes at leading order as~\cite{Boer1999,Arnold2009}
\begin{widetext}
\begin{eqnarray}
\frac{d\sigma}{dx_1 ~ dx_2 ~ d^2\bm{q}_T ~ d\Omega}&=&\frac{\alpha^2_{em}}{3Q^2} \Big\{ {(1 + \cos^2\theta)\over 4}F_{UU}^1 + S_{L} {\sin^2\theta\over 4}\sin2\phi F_{LU}^{\sin2\phi} \nonumber\\
 & &+ \lvert \bm{S}_{T}\rvert {\sin^2\theta\over 4}\left[\sin(2\phi + \phi_S)F_{TU}^{\sin(2\phi + \phi_S)} + \sin(2\phi - \phi_S)F_{TU}^{\sin(2\phi - \phi_S)} \right] + \cdots \Big\}. \label{dycs}
\end{eqnarray}
\end{widetext}
Here $\phi$ and $\phi_S$ are the azimuthal angles of $\bm{l}_\perp$
and $\bm{S}_{T}$ with respect to the normalized vector
$\bm{h}=\bm{q}_T/Q_T$, respectively; and $d\Omega=d\cos\theta d
\phi$ is the solid angle of the lepton $\ell$ in the center-of-mass
system of the lepton pair. In Eq.~(\ref{dycs}) we only give the
terms appearing in (\ref{ppdy}), and other terms do not contribute
in our analysis below. We note that in the literature there are
different definitions of angles $\phi$ and $\phi_S$, as shown in
Fig.~\ref{angledef}. In this work we adopt the definition in
Ref.~\cite{Arnold2009}. Also, we apply the so-called Collins-Soper
frame~\cite{CS_frame}, in which the structure functions are
expressed as
\begin{widetext}
\begin{eqnarray}
F_{UU}^{1}& =& \mathcal{C} \big[f_1\bar{f}_1\big],\\
F_{TU}^{\sin(2\phi - \phi_S)} &=& \mathcal{C} \Big[\frac{\bm{h}\cdot\bm{k}_{1T}}{M_N} h_1 \bar{h}_1^\perp\Big],
\label{eq:struc_trans_bm}\\
F_{TU}^{\sin(2\phi + \phi_S)} &=& \mathcal{C} \Big[\frac{2(\bm{h}\cdot\bm{k}_{1T})[2(\bm{h}\cdot\bm{k}_{1T})(\bm{h}\cdot\bm{k}_{2T}) - \bm{k}_{1T}\cdot\bm{k}_{2T}] - k_{1T}^2(\bm{h}\cdot\bm{k}_{2T})}{2M_N^3} h_{1T}^\perp \bar{h}_1^\perp\Big],
\label{eq:struc_pretz_bm}\\
F_{LU}^{\sin2\phi}& =& \mathcal{C} \Big[\frac{2(\bm{h}\cdot\bm{k}_{1T})(\bm{h}\cdot\bm{k}_{2T}) - \bm{k}_{1T}\cdot\bm{k}_{2T}}{M_N^2} h_{1L}^\perp \bar{h}_1^\perp\Big].
\label{eq:struc_h_bm}
\end{eqnarray}
\end{widetext}
In above equations we have used the notation
\begin{equation}
\begin{split}
\mathcal{C}\big[w(\bm{k}_{1T},\bm{k}_{2T})f\bar{g} \big] =&  \sum_q e_q^2 \int d^2 \bm{k}_{1T} ~ d^2 \bm{k}_{2T} ~ \\
&\times \delta^{(2)}\left(\bm{q}_T - \bm{k}_{1T} - \bm{k}_{2T} \right)  w(\bm{k}_{1T},\bm{k}_{2T}) \\
 & \times \left[ f^{q}( x_1 ,k_{1T}^2) g^{\bar{q}}( x_2 , k_{2T}^2)\right.
 \\
 &\left.+ f^{\bar{q}}( x_1 ,k_{1T}^2) g^q( x_2 , k_{2T}^2)\right].
\end{split}
\end{equation}
Thus all structure functions depend on $x_1$, $x_2$ and
$q_T=\lvert\bm{q}_T\rvert$.

As shown in Eq.~(\ref{dycs}), the structure functions $F_{TU}^{\sin(2\phi - \phi_S)}$ and $F_{TU}^{\sin(2\phi + \phi_S)}$ contribute to the cross section in the case in which one proton is transversely polarized (denoted by subscript $T$), and will give rise to $\sin(2\phi - \phi_S)$ and $\sin(2\phi + \phi_S)$ angular dependences, respectively. The structure function $F_{LU}^{\sin2\phi}$  contribute to the cross section in the case in which one proton is longitudinally polarized (denoted by the subscript $L$), and will give rise to a $\sin2\phi$ angular dependence. Therefore one can define the following azimuthal asymmetries
\begin{eqnarray}
A^{\sin(2\phi - \phi_S)}_{TU} (x_1,\,x_2,\,q_T) &=& {F_{TU}^{\sin(2\phi - \phi_S)} \over F_{UU}^1}, \label{asin2m}\\
A^{\sin(2\phi + \phi_S)}_{TU} (x_1,\,x_2,\,q_T) &=& {F_{TU}^{\sin(2\phi + \phi_S)} \over F_{UU}^1},\label{asin2p}\\
A^{\sin2\phi }_{LU} (x_1,\,x_2,\,q_T) &=& {F_{LU}^{\sin2\phi}\over F_{UU}^1}.\label{asin2}
\end{eqnarray}
Our definitions for the azimuthal asymmetries are similar to the
analyzing power given in \cite{Boer1999} and are different from the
transverse momentum weighted asymmetries defined in
\cite{Bacchetta:2010si}. For experimental measurement of the
asymmetries given in Eqs.(\ref{asin2m}), (\ref{asin2p}), and
(\ref{asin2}), the polar angle $\theta$ of the lepton $\ell$ should
be identified. As a compensation, larger asymmetries could be
measured.

One can also express the cross section of the Drell-Yan process, depending on $y$ and $Q^2$ as
\begin{equation}
\frac{d\sigma}{dy ~ dQ^2 ~ d^2\bm{q}_T ~ d\Omega} = \frac{1}{s} \frac{d\sigma}{dx_1 ~ dx_2 ~ d^2\bm{q}_T ~ d\Omega}.
\end{equation}
At the region $\bm{q}_T^2\ll Q^2$, the following relations hold
\begin{equation}
x_1 = \frac{Q}{\sqrt{s}}e^y, ~~~x_2 = \frac{Q}{\sqrt{s}}e^{-y}.
\label{eq:xdef}
\end{equation}

Therefore we can define
the $y$-dependent and $Q^2$-dependent SSAs as
\begin{eqnarray}
A_{PU}^a(y) &=& \frac{\int d^2\bm{q}_T~dQ^2 \frac{1}{Q^2}F_{PU}^a}{\int d^2\bm{q}_T ~dQ^2 \frac{1}{Q^2}F_{UU}^1}, \label{apuy} \\ A_{PU}^a(Q^2) &=& \frac{\int d^2\bm{q}_T~dy F_{PU}^a}{\int d^2\bm{q}_T~dy F_{UU}^1},\label{apuq}
\end{eqnarray}
where we have used the short notes $P=T$ or $L$, and $a=\sin(2\phi
\pm \phi_S) $ for $P=T$ and $a=\sin2\phi$ for $P=L$. The
integrations in Eqs.(\ref{apuy}) and (\ref{apuq}) are performed
according to kinematical cuts or experimental acceptances.

\section{Phenomenological analysis of azimuthal asymmetries at RHIC, J-PARC, E906 and NICA}

In this section we investigate the prospects of experimental
measurements on the azimuthal asymmetries defined in the last
section at various facilities that can conduct single polarized
proton-proton Drell-Yan processes. The proton-proton Drell-Yan
process involves the annihilation of a quark from one proton and a
antiquark from another proton. In order to calculate $F_{PU}^a $,
one needs to know the distributions $h_1$, $h_{1T}^\perp$,
$h_{1L}^\perp$, and $h_1^\perp $ of both the valence and sea quarks.
Although there are some extractions of transversity and Boer-Mulders
functions from SIDIS and Drell-Yan data, most of the chiral-odd
parton distributions are not measured and less known, especially
those of sea quarks. In order to estimate the azimuthal asymmetries
in $pp$ Drell-Yan processes, we apply the following ansatz:

\begin{figure*}[t]
\begin{center}
\includegraphics[width=0.32\textwidth]{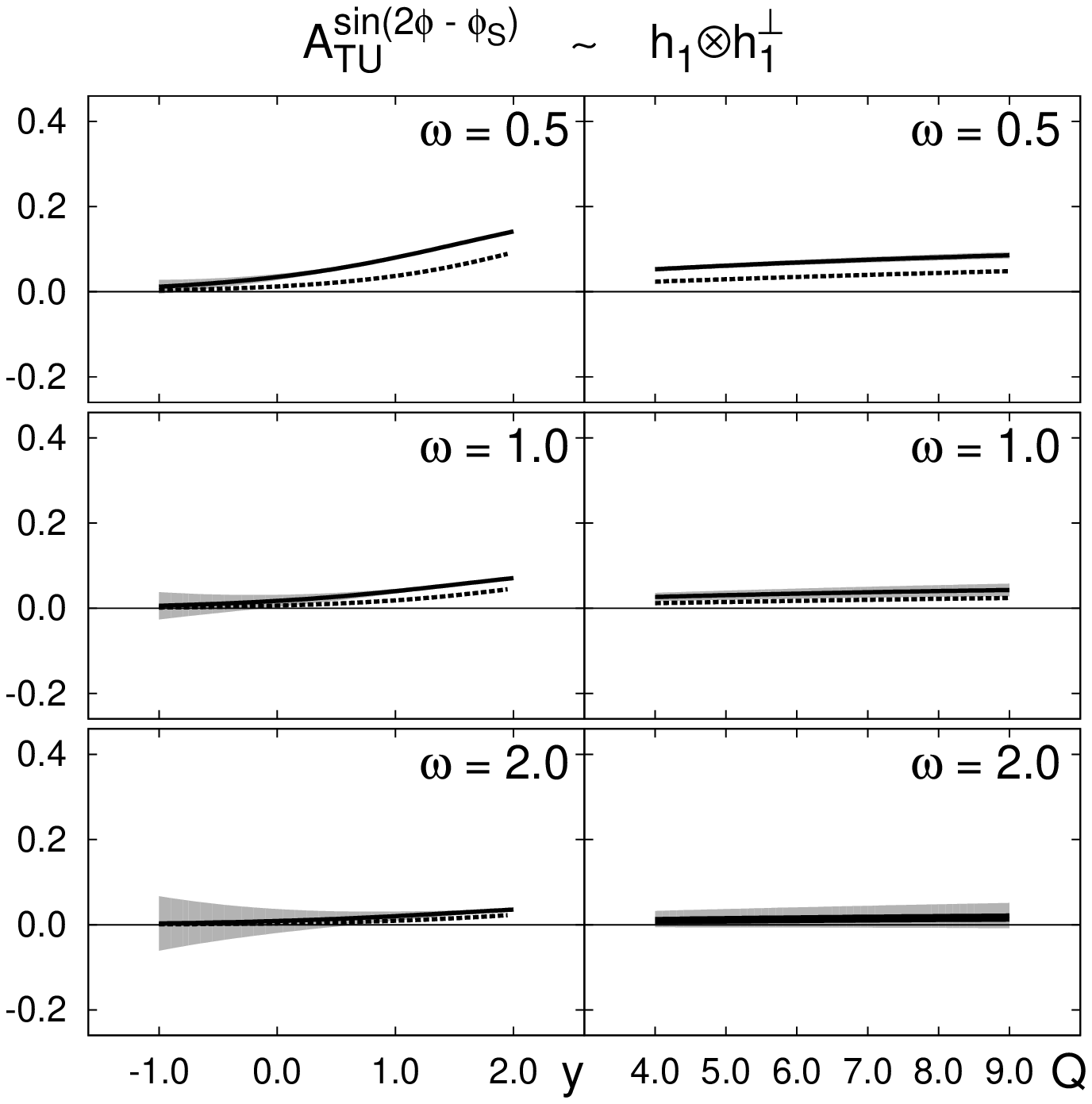}
\includegraphics[width=0.32\textwidth]{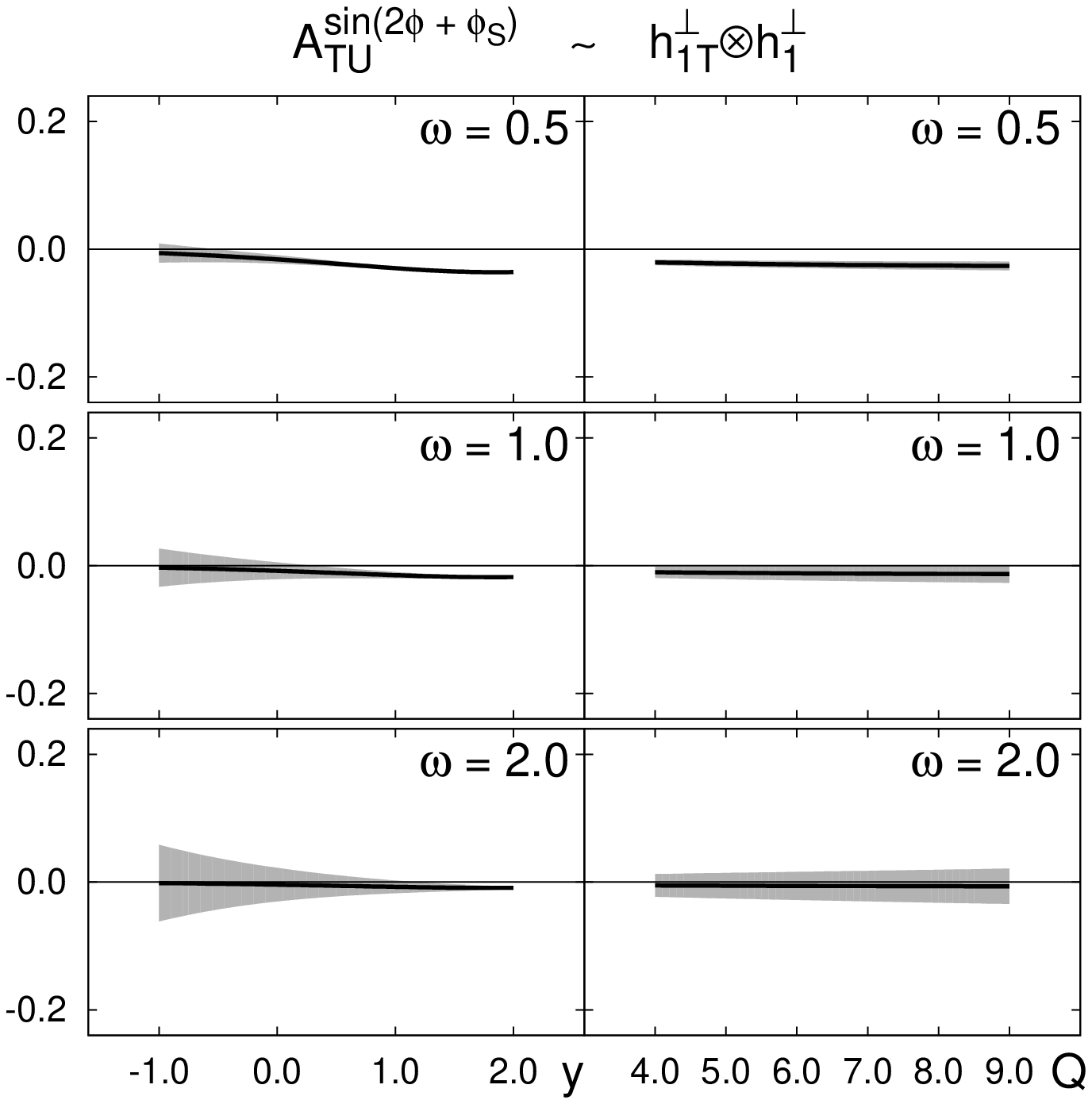}
\includegraphics[width=0.32\textwidth]{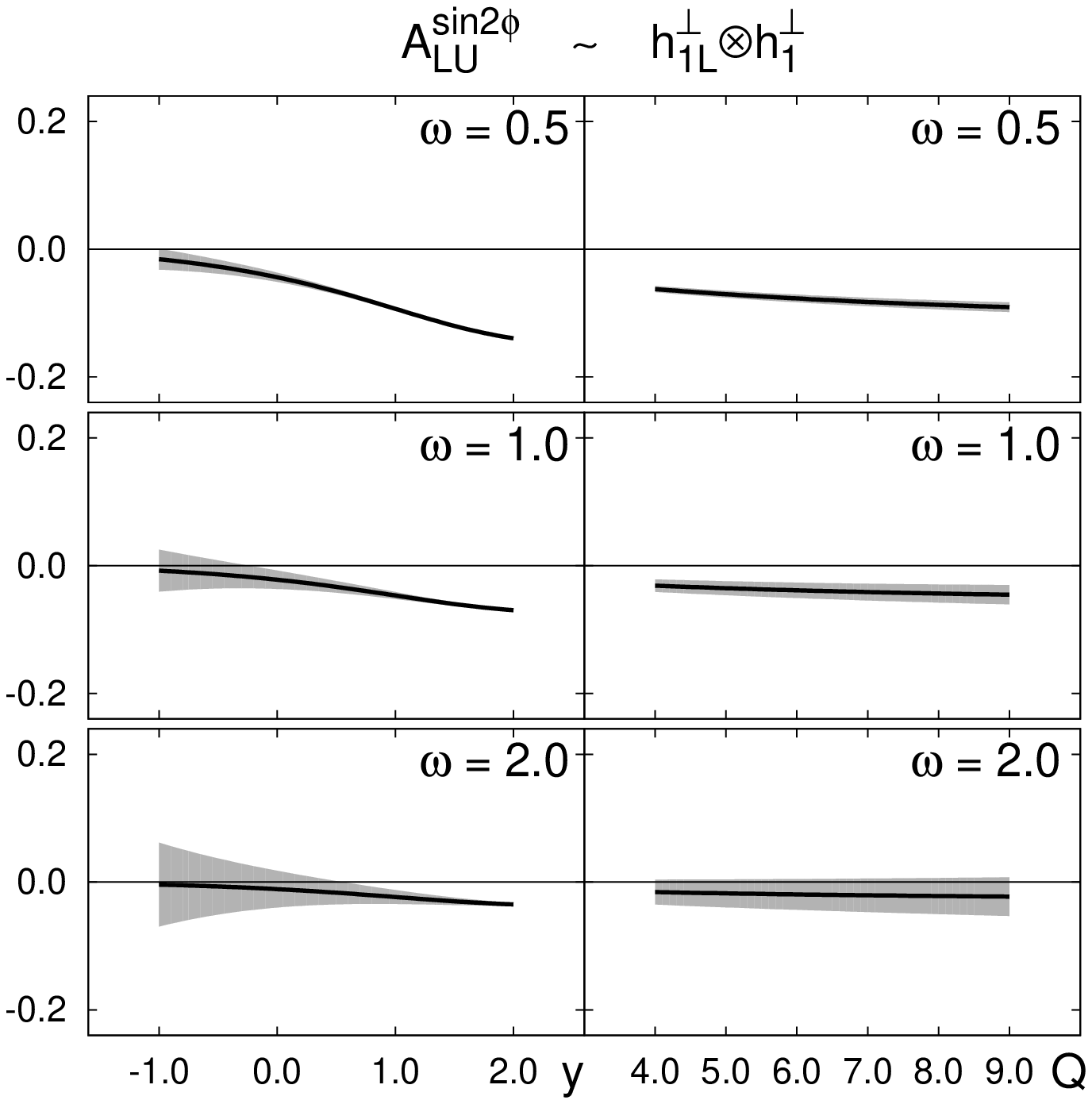}
\caption{ Azimuthal asymmetries $A_{TU}^{\sin(2\phi-\phi_S)}$
(left panels), $A_{TU}^{\sin(2\phi+\phi_S)}$ (central panels), and
$A_{TU}^{\sin2\phi}$ (right panels) at RHIC collider experiments as
functions of the rapidity $y$ and the dilepton mass $Q$,
respectively. The dashed lines in the left panels correspond to the
contributions from the valence transversity
distributions $h_1$ fitted by Anselmino \textit{et al.} in
Ref.~\cite{Anselmino:2008jk}. The thick solid lines in the left, central, and right panels represent
the contributions from the distributions $h_1$, $h_{1T}^\perp$, and
$h_{1L}^\perp$ of valence quarks alone in the light-cone quark-diquark
model~\cite{Ma,Schmidt1997,Shejun2009,Zhu2011}.
The shaded regions give the ranges of the asymmetries
by considering the additional contribution from the distributions
$h_1$, $h_{1T}^\perp$, and $h_{1L}^\perp$ of sea quarks constrained
by the positivity bounds given in Eqs. (\ref{h1bound}),
(\ref{pretzbound}) and (\ref{hbound}). The upper and lower limits of
the bands correspond to the asymmetries by saturating the positivity
bounds.}\label{fig:rhic_collider}
\end{center}
\end{figure*}

\begin{itemize}
\item For the Boer-Mulders functions $h_1^{\perp q}(x, k_T^2)$, we adopt the result extracted from
the unpolarized $pd$~\cite{e866} and $pp$~\cite{e866pp} Drell-Yan data in Ref.~\cite{Lu2009}, as there is parameterization for both valence and sea quarks with the following
form:
\begin{eqnarray}
h_1^{\perp q}(x, k_T^2) &=& H_qx^{c^q}(1-x)^{b}f_1^q(x)\nonumber\\
&&\times\frac{1}{\pi k_{bm}^2}\exp\bigg(\frac{-k_T^2 }{ k_{bm}^2}\bigg),
\end{eqnarray}
where the subscript ``$bm$" stands for the Boer-Mulders functions, and $q=u,\,d,\,\bar{u}$, and $\bar{d}$.
We have ignored the
contributions from other flavors, since they are assumed to be
small. We note that the possible range of parameters $H_q$ allowed
by the positivity bound for $h_1^\perp$ can be described by the
coefficient $\omega$, namely, that the substitutions $H_q\to\omega
H_q$ for $q = u,\, d$ and $H_q\to\frac{1}{\omega}H_q$ for $q =
\bar{u},\, \bar{d}$ will not change the calculated $\cos2\phi$
asymmetry (contributed by $h_1^{\perp q}\times h_1^{\perp \bar{q}}$)
in the unpolarized $pd$ and $pp$ Drell-Yan data.  The range of
$\omega$ given in Ref.~\cite{Lu2009} is $0.48 < \omega < 2.1$ and
$\omega=1$ corresponds to the central values of $H_q$. However, for
the azimuthal asymmetries given in Eqs.~(\ref{apuy}) and
(\ref{apuq}), the variation of $\omega$ will lead to the change of
the magnitudes of the asymmetries, and will be considered in our
calculations.

\item
For the $T$-even distributions $h_1$, $h_{1T}^\perp$, and
$h_{1L}^\perp$ of valence quarks, there are considerable model
calculations. We will deploy the calculation from the
light-cone quark-diquark model. In this model, the Melosh-Wigner
rotation~\cite{Wigner}, which plays an important role to understand
the proton spin puzzle~\cite{MaOld} due to the relativistic effect
of quark transversal motions, has been taken into account. In
practice, the light-cone quark-diquark model has been applied to
calculate the helicity distributions~\cite{Ma:1996sr}, the
transversity distributions~\cite{Schmidt1997,Ma} and other
3dPDFs~\cite{Lu:2004au,Shejun2009,Zhu2011}, and related azimuthal
spin asymmetries in SIDIS processes~\cite{Ma:2001ie,Ma:2002ns}.

The light-cone model results for the distributions $h_1^{qv}(x, k_T^2)$, $h_{1T}^{\perp qv}(x, k_T^2)$, and $h_{1L}^{\perp qv}(x, k_T^2)$ are given as~\cite{Ma,Schmidt1997,Shejun2009,Zhu2011}
\begin{align}
j^{uv}(x, k_T^2) =& \big[f_1^{uv}(x, k_T^2) - \frac{1}{2} f_1^{dv}(x, k_T^2)\big] W_S^j(x, k_T^2) \nonumber\\
&-\frac{1}{6} f_1^{dv}(x, k_T^2) W_V^j(x, k_T^2),\\
j^{dv}(x, k_T^2) =& -\frac{1}{3} f_1^{dv}(x, k_T^2) W_V^j(x, k_T^2),
\label{eq:tevenpdf}
\end{align}
where $j = h_1, \,h_{1T}^\perp,\, h_{1L}^\perp$, respectively, and the superscript ``$v$" is corresponding to the valence distributions. $W_{S/V}^j(x, k_T^2)$ are the rotation factors for the scalar or axial vector spectator-diquark cases. Their explicit form are
\begin{eqnarray}
W_D^{h_1}(x, k_T^2) = \frac{(x\mathcal{M}_D + m_q)^2}{(x\mathcal{M}_D + m_q)^2 + k_T^2},\\
W_D^{h_{1T}^\perp}(x, k_T^2) = -\frac{2M_N^2}{(x\mathcal{M}_D + m_q)^2 + k_T^2},\\
W_D^{h_{1L}^\perp}(x, k_T^2) = -\frac{2M_N(x\mathcal{M}_D + m_q)}{(x\mathcal{M}_D + m_q)^2 + k_T^2},
\end{eqnarray}
with
\begin{equation}
\mathcal{M}_D = \sqrt{\frac{m_q^2 + k_T^2}{x} + \frac{m_D^2 + k_T^2}{1- x}}.
\end{equation}

\begin{figure*}[t]
\begin{center}
\includegraphics[width=0.32\textwidth]{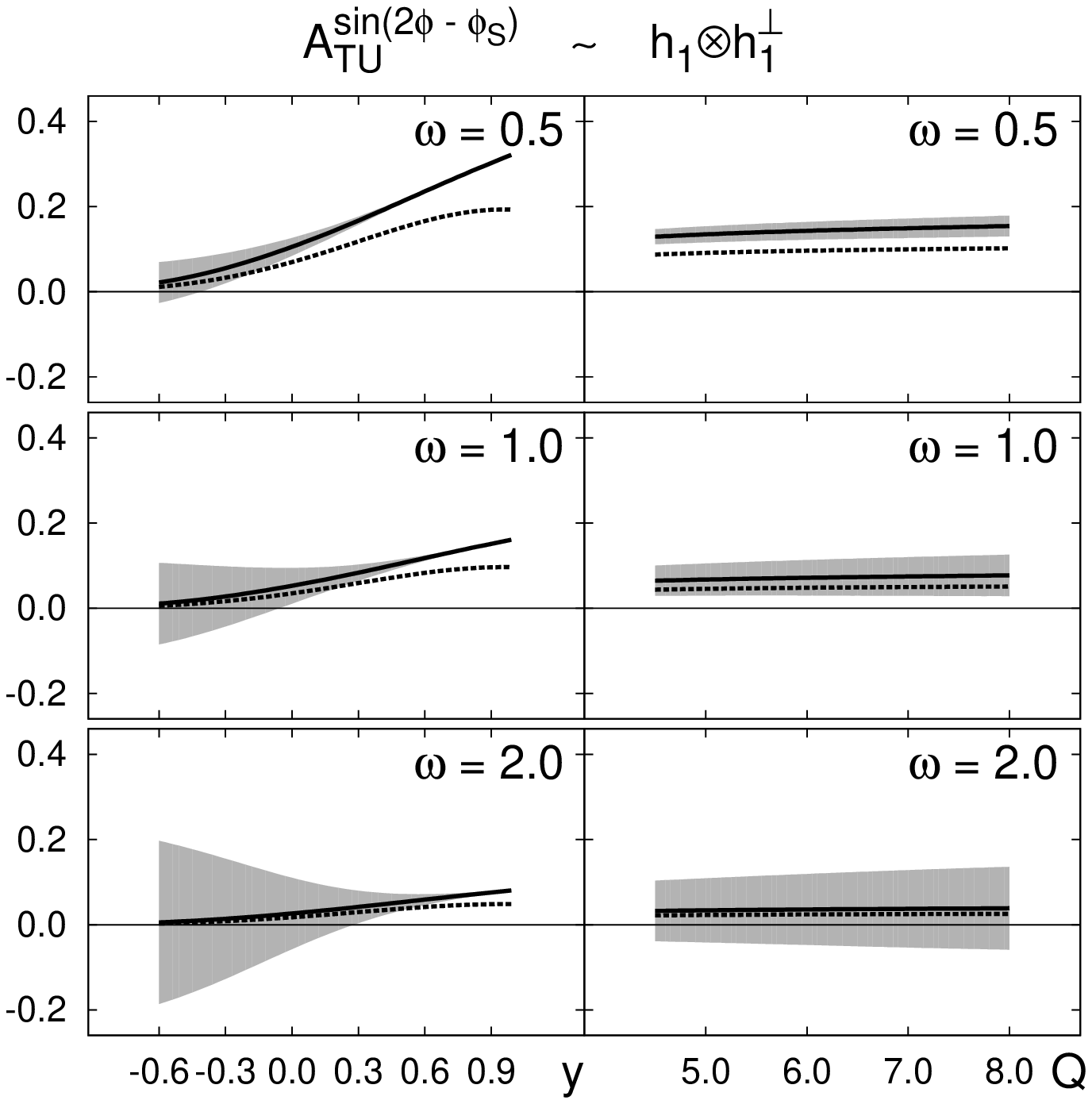}
\includegraphics[width=0.32\textwidth]{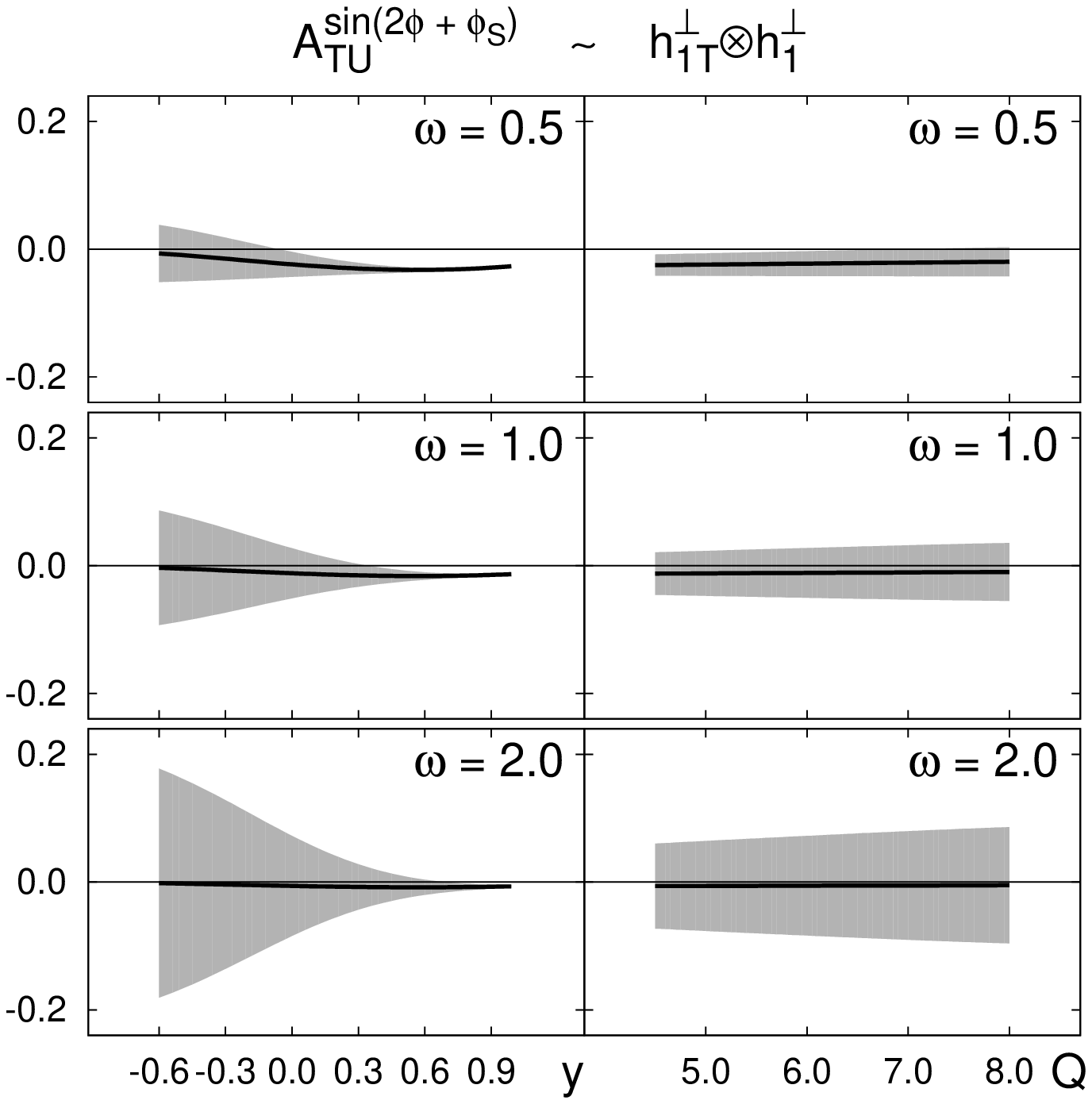}
\includegraphics[width=0.32\textwidth]{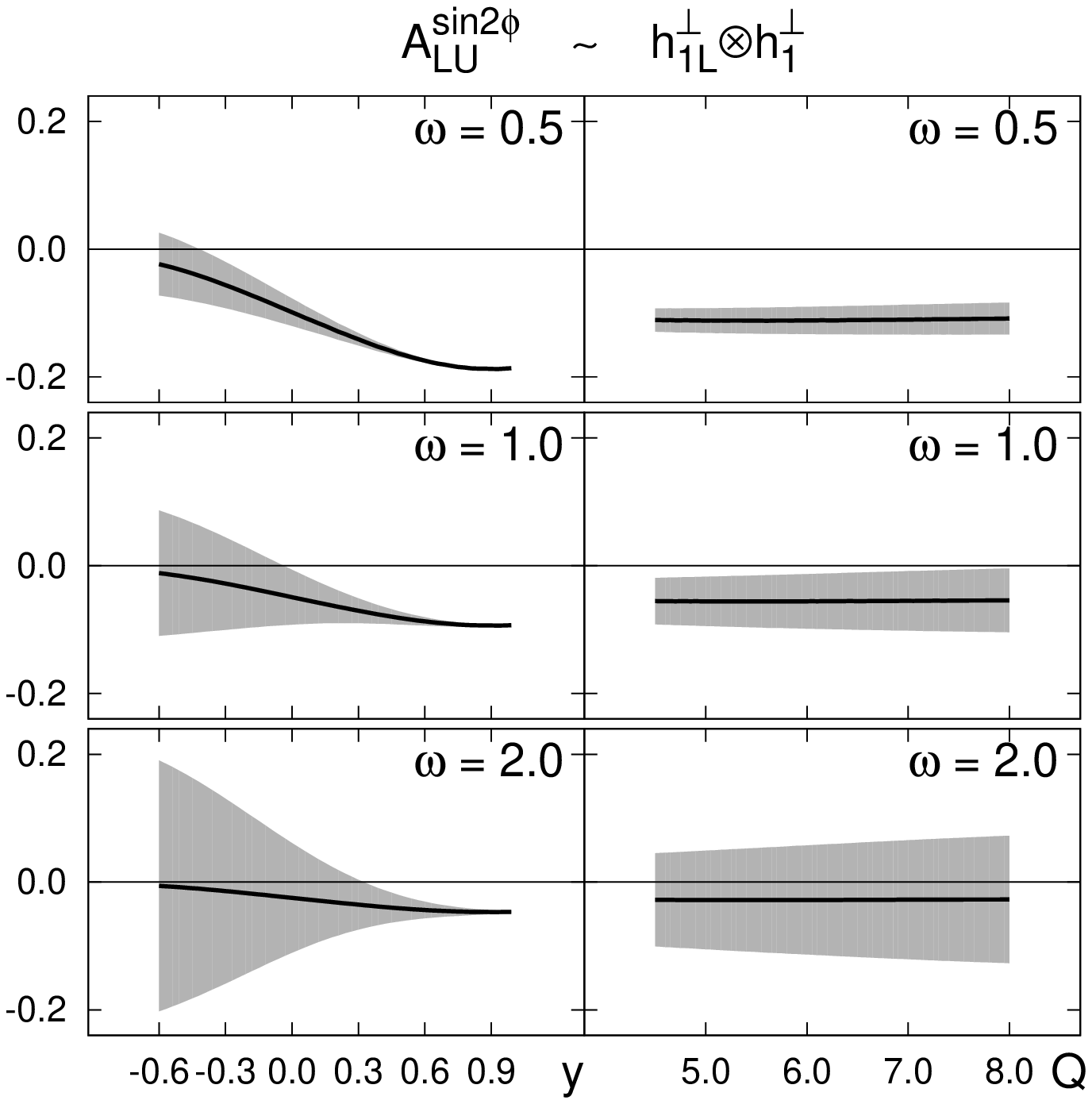}
\caption{\label{fig:rhic_fixed} Azimuthal asymmetries $A_{TU}^{\sin(2\phi-\phi_S)}$ (left panels) , $A_{TU}^{\sin(2\phi+\phi_S)}$ (central panels),  and $A_{LU}^{\sin2\phi}$ (right panels)  at RHIC fixed-target experiments.}
\end{center}
\end{figure*}

An important feature manifested by these rotation factors is that they automatically satisfy the requirement of the positivity bounds~\cite{Bacchetta2000} for the PDFs. In the left, central, and right panels of Fig.~\ref{fig:teventmd} we plot the curves for $x h_1(x)$, $x h_{1T}^{\perp(1)}(x)$, and $x h_{1T}^{\perp(1)}(x)$ of valence $u$ and $d$ quarks at $Q^2=1$ $\mathrm{GeV}^2$, respectively.

As there is already extraction of transversity from the global analysis by combining
the SIDIS and $e^+e^-$ annihilation data, we also use the most recent
parametrizations~\cite{Anselmino:2008jk} for $h_1$ to calculate the asymmetries
$A_{TU}^{\sin2\phi}$, and compare the results with those predicted from our model calculation.

\item In order to consider the effects of the distributions $h_1$, $h_{1T}^\perp$ and $h_{1L}^\perp$ of sea quarks, we constrain them by the positivity
bounds~\cite{Bacchetta2000}
\begin{eqnarray}
\Big\lvert h_1^{\bar{q}} (x,\,k_T^2) \Big\rvert \leqslant f_1^{\bar{q}}(x,\,k_T^2),\label{h1bound}\\
\Big\lvert \frac{k_T^2}{2M_N^2}h_{1T}^{\perp \bar{q}}(x,\,k_T^2) \Big\rvert \leqslant f_1^{\bar{q}}(x,\,k_T^2), \label{pretzbound}\\
 \Big\lvert \frac{k_T}{M_N}h_{1L}^{\perp \bar{q}} (x,\,k_T^2)\Big\rvert \leqslant f_1^{\bar{q}}(x,\,k_T^2).\label{hbound}
\end{eqnarray}
They give rise to additional contributions to the asymmetries through
the coupling $j^{\bar{q}}\otimes h_1^{\perp q}$, and will give a
range of the asymmetries by varying the distributions within the
bounds. By saturating the positivity bounds, one can obtain the
upper and lower limits of the asymmetries.

\item  For the unpolarized distributions $f_1^q(x, k_T^2)$, we use the MSTW2008 LO set parametrization~\cite{Martin2009}, and adopt a Gaussian form factor for the transverse momentum dependence which has been adopted in many phenomenological analyses~\cite{Anselmino:2007fs,Anselmino:2008jk}
\begin{equation}
f_1^q(x, k_T^2) = f_1^q(x)\frac{\exp(-k_T^2 / k_{un}^2)}{\pi k_{un}^2},
\end{equation}
with $k_{un}^2 = 0.25~\mathrm{GeV}^2$; the subscript ``$un$" stands for the unpolarized distributions.

\item In order to precisely predict the azimuthal asymmetries at different 
experiments using TMD
factorization, it is essential for one to know the evolution of 3dPDFs.
Unlike the PDFs in the collinear factorization approach, whose evolution has been 
well established by the DGLAP 
equation, the $Q^2$-dependence of 3dPDF is not fully understood yet.
In our practical calculations we assume that the scale dependences of 3dPDF and 
the spin averaged distribution function $f_1$ are the same.
The same assumption has been applied in some extractions of 
3dPDF~\cite{Zhang:2008nu,Anselmino:2008jk,Anselmino:2008sga,Lu2009}.
To what extent that this approximation is valid still needs to be studied.
As the asymmetries we calculate are ratios, we expect that our assumption
on the scale dependence are reasonable.

\end{itemize}

Now we have all the ingredients for estimating the azimuthal asymmetries in single polarized proton-proton Drell-Yan processes. In the following, we apply the above ansatz to present our predictions and phenomenological analysis for forthcoming experiments at RHIC, J-PARC, E906, and NICA.

\begin{itemize}
\item RHIC

The original proposal of Drell-Yan experiment at RHIC employs two
proton beams to collide at $\sqrt{s}=200~\mathrm{GeV}$ or
$500~\mathrm{GeV}$~\cite{Bunce:2000uv}. But recently there is also a
new proposal to conduct a fixed-target experiment at
$\sqrt{s}=22~\mathrm{GeV}$~\cite{Goto:2011zz}. We estimate the
asymmetries for both the collider and fixed-target modes at RHIC.
The longitudinal proton beam will be run in the coming years at
RHIC, after that, Drell-Yan program with tranverse spin will be
conducted. We choose the following kinematics for collider experiment at
RHIC-STAR (Solenoidal Tracker at RHIC):
 \[ \begin{split}
&\sqrt{s} = 200~\mathrm{GeV},~~4~\mathrm{GeV}<Q<9~\mathrm{GeV}, \\
&0<q_T < 1~\mathrm{GeV},~~-1 < y < 2.
\end{split}\]
We constrain the kinematical cut at the low transverse momentum region such that $q_T^2\ll Q^2$ where TMD factorization dominates.

\begin{figure*}[t]
\begin{center}
\includegraphics[width=0.32\textwidth]{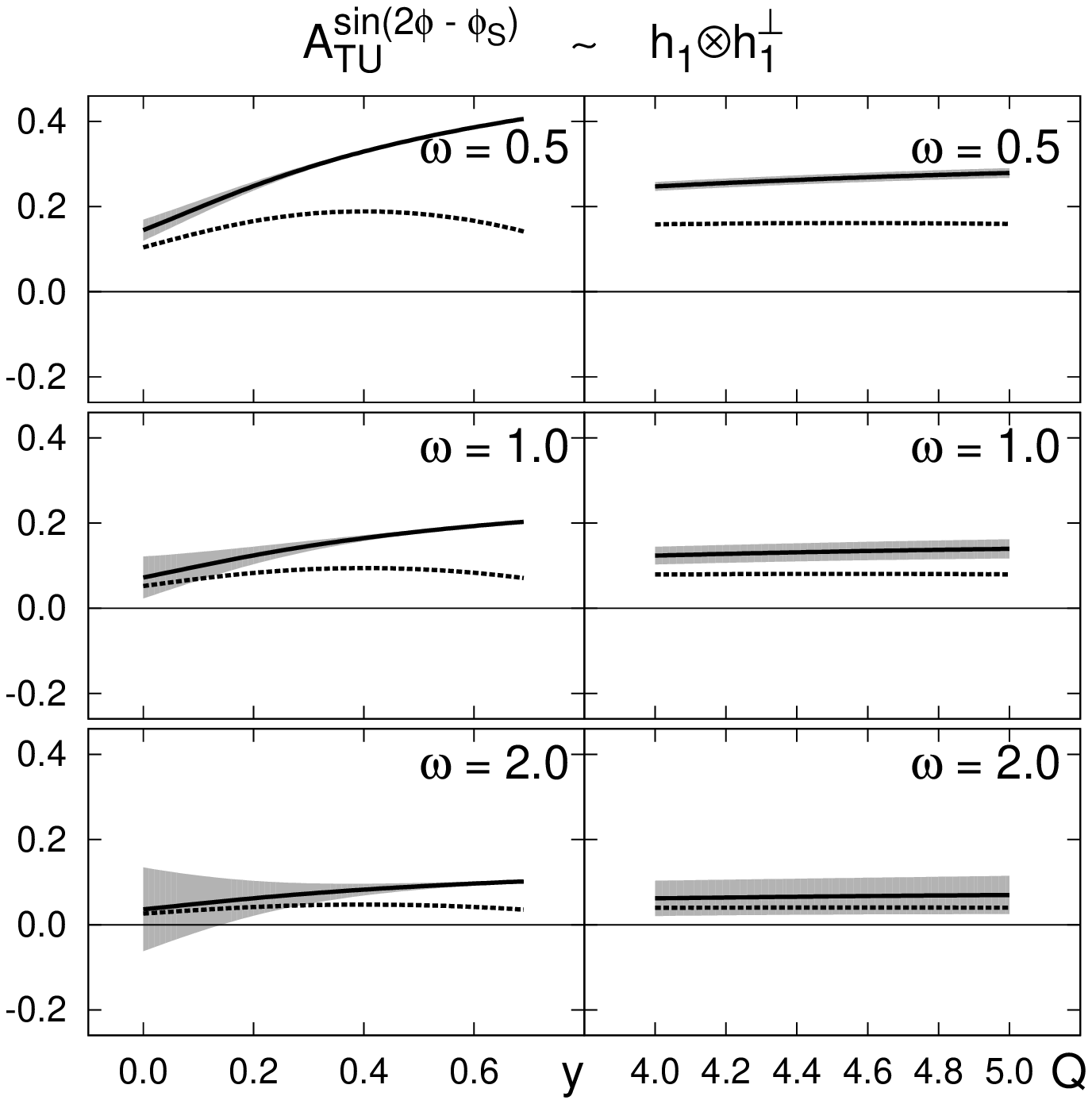}
\includegraphics[width=0.32\textwidth]{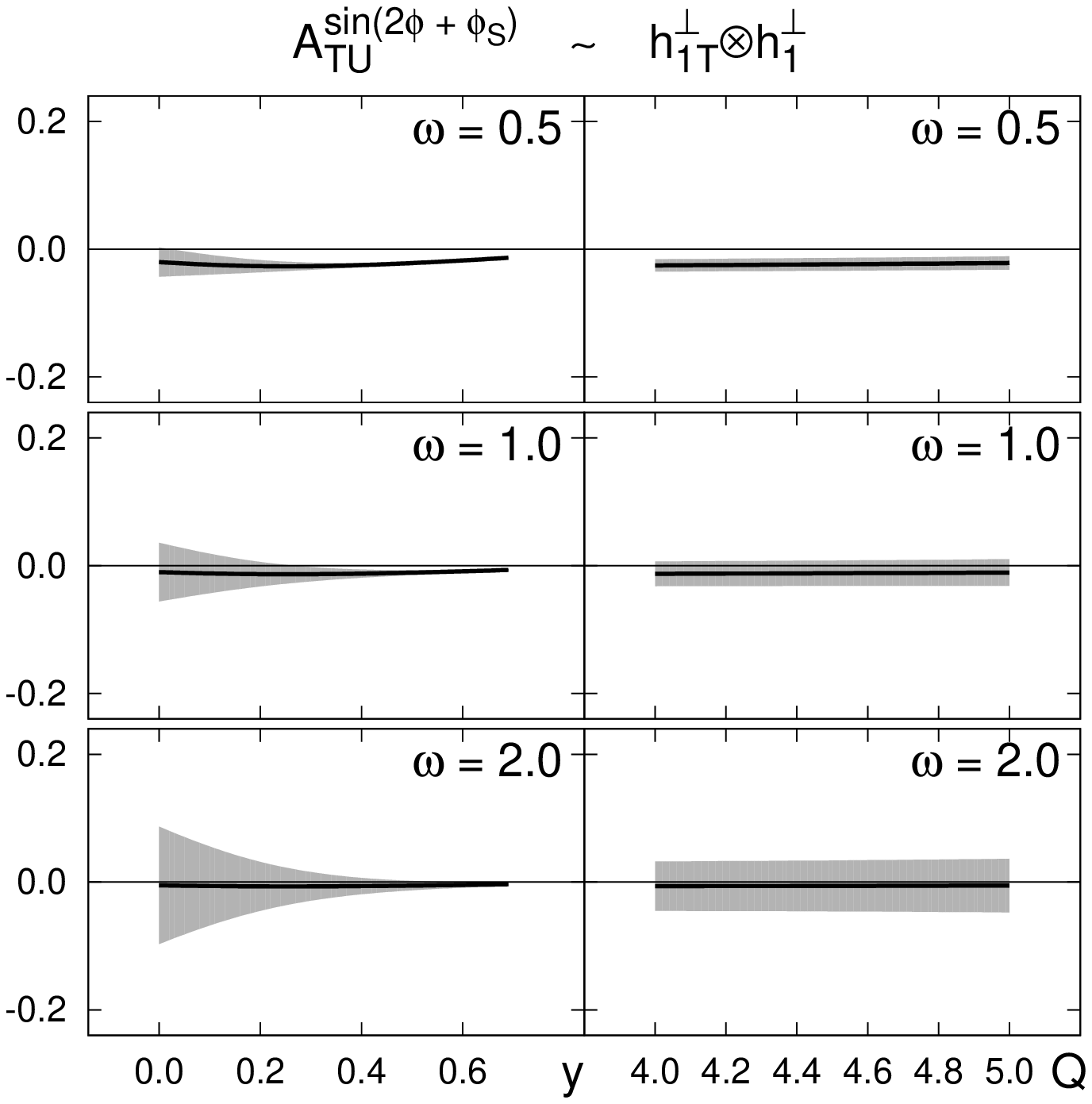}
\includegraphics[width=0.32\textwidth]{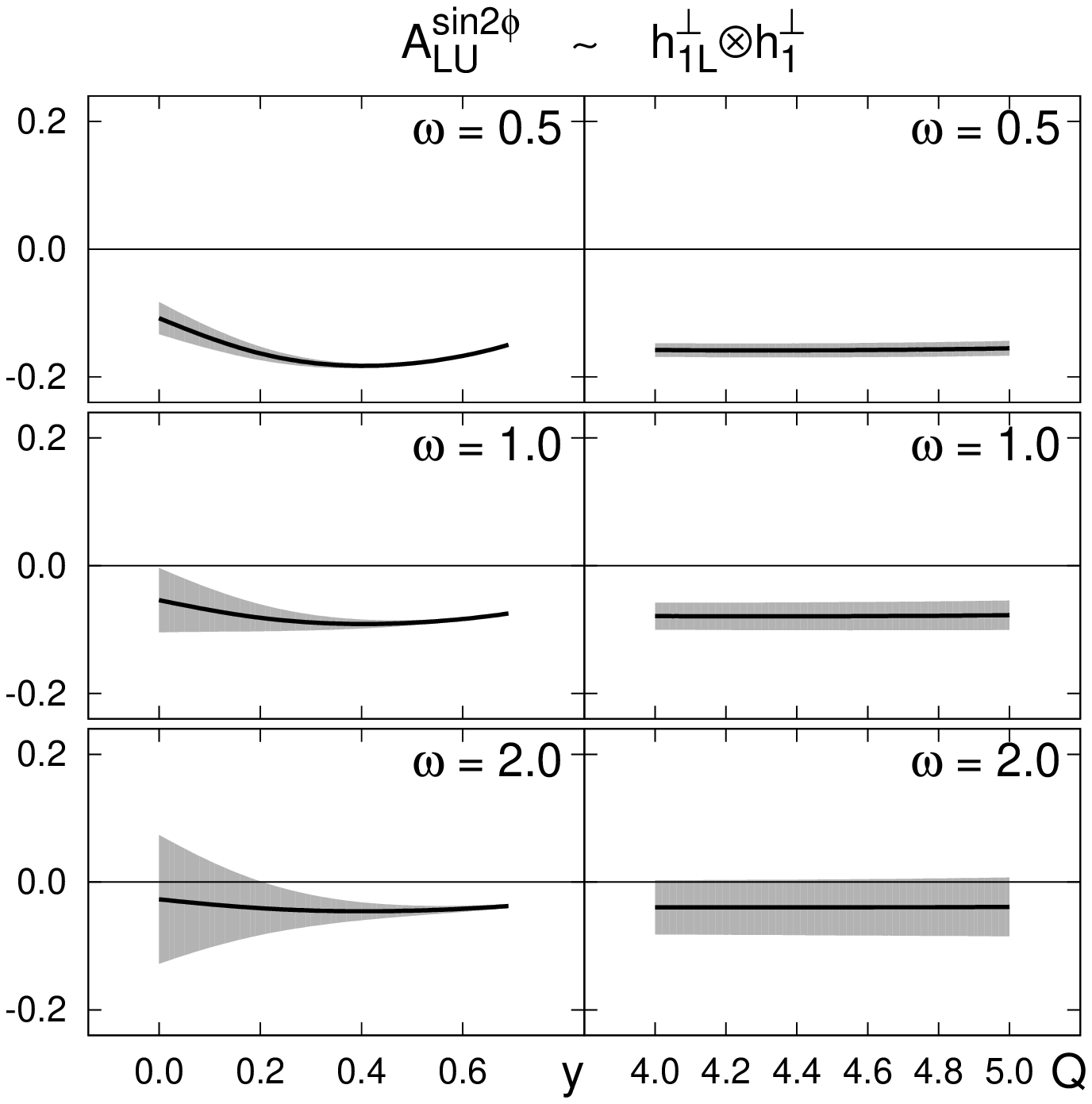}
\caption{ Azimuthal asymmetries $A_{TU}^{\sin(2\phi-\phi_S)}$ (left panels) , $A_{TU}^{\sin(2\phi+\phi_S)}$ (central panels),  and $A_{LU}^{\sin2\phi}$ (right panels)  at J-PARC.}\label{fig:jparc}
\end{center}
\end{figure*}

In the left panels of Fig.~\ref{fig:rhic_collider} we show the
estimated azimuthal asymmetries $A_{TU}^{\sin(2\phi-\phi_S)}$ as
functions of the rapidity $y$ and the dilepton mass $Q$,
respectively. The difference between two linestyles in the left panels is that
for the dashed lines we use the transversity distribution for valence $u$ and $d$ quarks
from the parameterization in Ref.~\cite{Anselmino:2008sga},
while for the solid lines we adopt the results from
the light-cone quark-diquark model~\cite{Ma,Schmidt1997}
for the valence distributions $h_1$.
The thick solid and dashed lines correspond to the contribution
merely from the combinations of the valence transversity
distributions and the sea Boer-Mulders distributions, that is,
ignoring the transversity distributions of sea quarks. The shaded
regions give the ranges of $A_{TU}^{\sin(2\phi - \phi_S)}$ by
considering the additional contribution of the transversity
distributions of sea quarks constrained by positivity bound
(\ref{h1bound}). The upper and lower limits of the bands correspond
to the asymmetries by saturating the positivity bound. The first,
second, and third rows show the results for $\omega=0.5$, $1$ and $2$,
respectively, where $\omega$ is the parameter for Boer-Mulders
functions, as explained previously.
In the central and right panels of Fig.~\ref{fig:rhic_collider} we show the estimated azimuthal
asymmetries $A_{TU}^{\sin(2\phi+\phi_S)}$ and $A_{LU}^{\sin2\phi}$,
respectively, in the same way as in the left panels.

From Fig.~\ref{fig:rhic_collider} we observe that in the forward
rapidity region, the asymmetry $A_{TU}^{\sin(2\phi-\phi_S)}$ at RHIC
collider experiments is positive, while the asymmetries
$A_{TU}^{\sin(2\phi+\phi_S)}$ and $A_{LU}^{\sin2\phi}$ tend to be
negative.  It is interesting that the magnitudes of the asymmetries
increase as the rapidity increases. At large forward rapidity, the
asymmetry $A_{TU}^{\sin(2\phi-\phi_S)}$ is dominated by the combination of the transversity of
valence quarks and Boer-Mulders function of sea quarks (showed by
the thick solid lines). This is understandable since large $y$
corresponds to larger $x_1$ and smaller $x_2$.  Therefore the
measurement of the asymmetries at large rapidity can provide the
information of $T$-even chiral-odd  distributions in valence region.
Of course, the statistics at large rapidity are much lower than
at midrapidity, therefore a reliable measurement requires data with
high integrated luminosity. A common feature shared by all these
asymmetries is that as $\omega$ increases, the asymmetry in the
forward rapidity region tends to decrease. This arises from the fact
that larger $\omega$ corresponds to larger valence Boer-Mulders
function and smaller sea Boer-Mulders function.

In the left panels of Fig.~\ref{fig:rhic_collider}, the magnitudes of the
asymmetries $A_{TU}^{\sin(2\phi-\phi_S)}$ calculated by using two different
forms of the valence transversity distributions $h_1$ are quite different at large
rapidity region. This is due to the fact that at large $x$ region the valence
transversity distributions $h_1$ fitted by Anselmino \textit{et al.} in
Ref.~\cite{Anselmino:2008jk}  are smaller than the corresponding ones in
the light-cone quark-diquark model~\cite{Ma,Schmidt1997}.

There is also the possibility of accelerating the polarized proton beam
with $E_p=250~\mathrm{GeV}$ to collide on the proton target at RHIC.
The RHIC kinematics for the fixed-target experiment are
\[\begin{split}
&\sqrt{s} = 22~\mathrm{GeV},~~4.5~\mathrm{GeV}<Q<8~\mathrm{GeV}, \\
&0<q_T < 1~\mathrm{GeV},~~0.2 <x_1 <0.6,
\end{split}\]
corresponding to $-0.6 < y < 1.0$, which is complementary to the
collider kinematics. In Fig.~\ref{fig:rhic_fixed} we show the
azimuthal asymmetries  $A_{TU}^{\sin(2\phi-\phi_S)}$ (left panels),
$A_{TU}^{\sin(2\phi+\phi_S)}$ (central panels), and
$A_{LU}^{\sin2\phi}$ (right panels) at RHIC fixed-target Drell-Yan
processes. It seems that the magnitude of the asymmetries
$A_{TU}^{\sin(2\phi-\phi_S)}$ and $A_{LU}^{\sin2\phi}$ at
fixed-target experiments are larger than that at collider
experiments. Therefore, there is a good chance to measure larger
asymmetries at the fixed-target mode. The drawback is that, at
fixed-target experiments, the uncertainty from the $T$-even
chiral-odd distributions $h_1$, $h_{1T}^\perp$, and $h_{1L}^\perp$ of
sea quarks at the negative rapidity region is larger than that at
collider experiments. In both modes the asymmetries are consistent
to zero at the large backward region and their size increases
with the increase of the rapidity.

\item J-PARC

J-PARC might measure azimuthal asymmetries given in Eqs.~(\ref{apuy},\ref{apuq}) in single polarized Drell-Yan processes at $E_p=50~\mathrm{GeV}$~\cite{Goto:2010zz}, corresponding
to $\sqrt{s} \simeq 10~\mathrm{GeV}$. The kinematical cuts at J-PARC are
\[\begin{split}
&4~\mathrm{GeV}<Q<5~\mathrm{GeV},~~0<q_T < 1~\mathrm{GeV}, \\
&0.5<x_1<0.9,
\end{split}\]
corresponding to $0<y<0.69$.
The estimated asymmetries  $A_{TU}^{\sin(2\phi-\phi_S)}$, $A_{TU}^{\sin(2\phi+\phi_S)}$, and $A_{LU}^{\sin2\phi}$ are shown in the left, central, and right panels of Fig.~\ref{fig:jparc}. The figure manifests that the asymmetry
$A_{TU}^{\sin(2\phi-\phi_S)}$ is positive, while the asymmetry $A_{LU}^{\sin2\phi}$ is negative in all the allowed rapidity region. This feature can be seen even by considering the uncertainty at the midrapidity region from the distributions $h_1$ and $h_{1L}^\perp$ of sea quarks. Larger asymmetries $A_{TU}^{\sin(2\phi-\phi_S)}$ and $A_{LU}^{\sin2\phi}$ are predicted at J-PARC than those at RHIC. The asymmetry $A_{TU}^{\sin(2\phi+\phi_S)}$ is much smaller than other two asymmetries, as at RHIC.

\begin{figure*}[t]
\begin{center}
\includegraphics[width=0.32\textwidth]{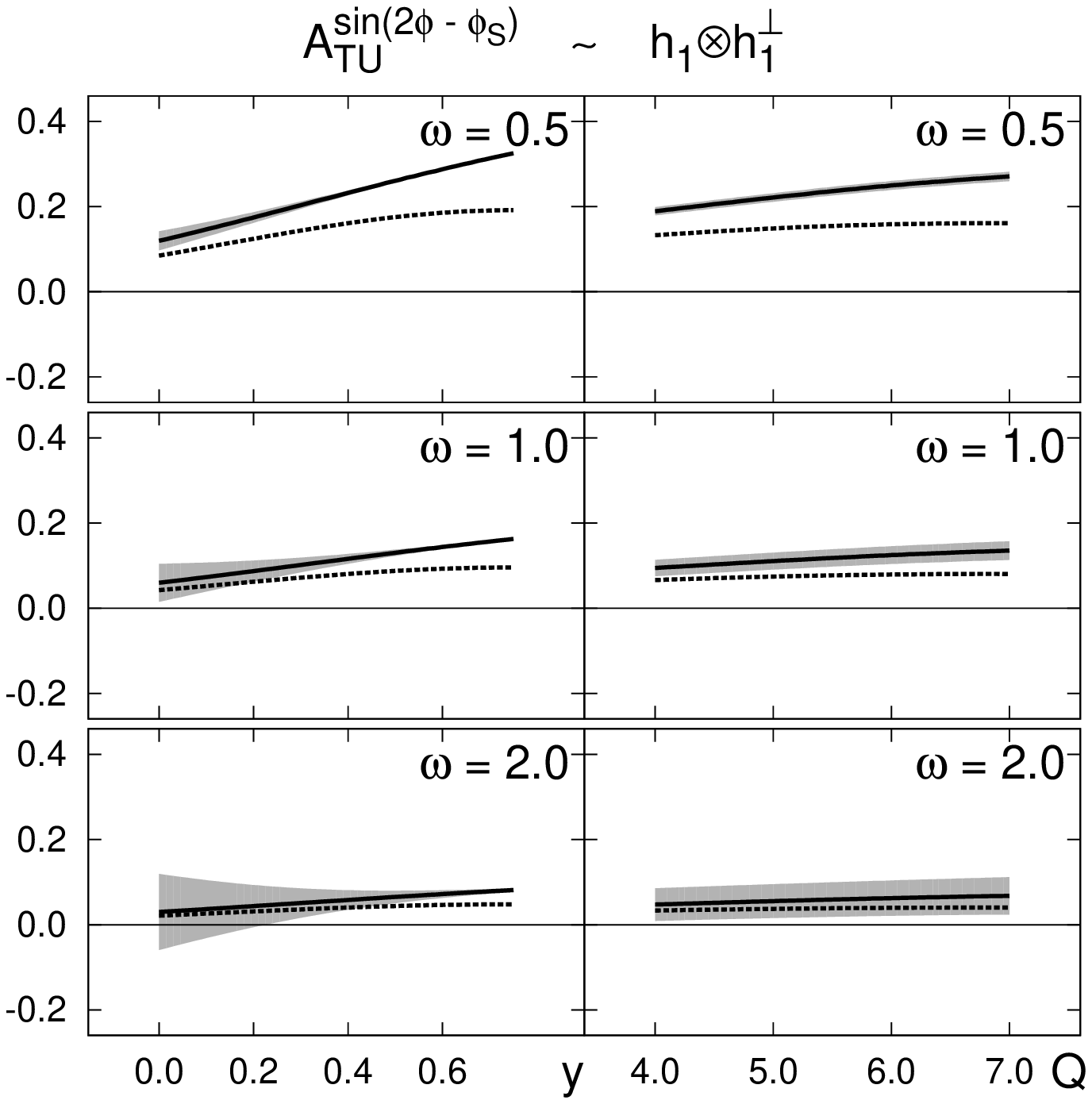}
\includegraphics[width=0.32\textwidth]{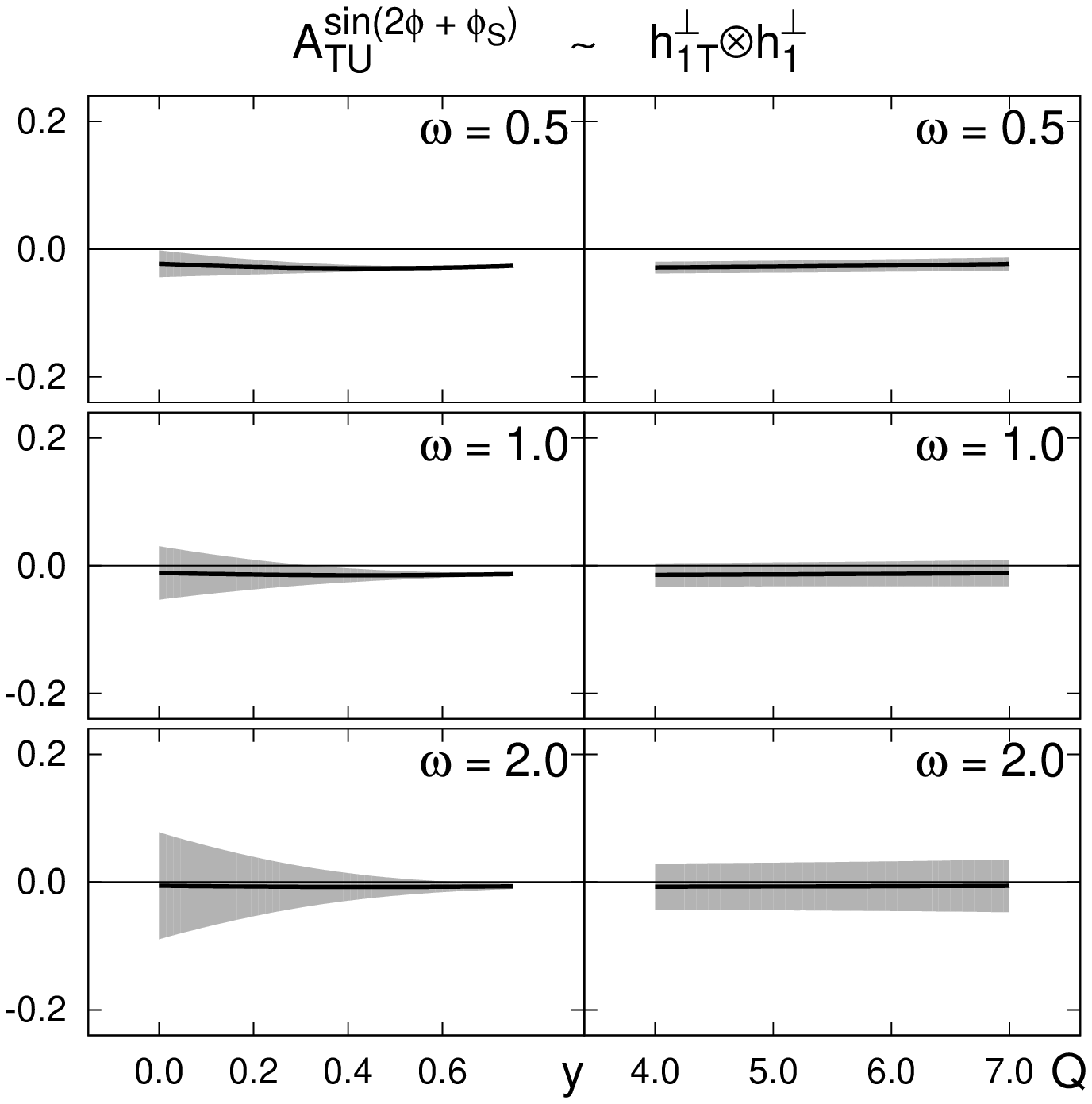}
\includegraphics[width=0.32\textwidth]{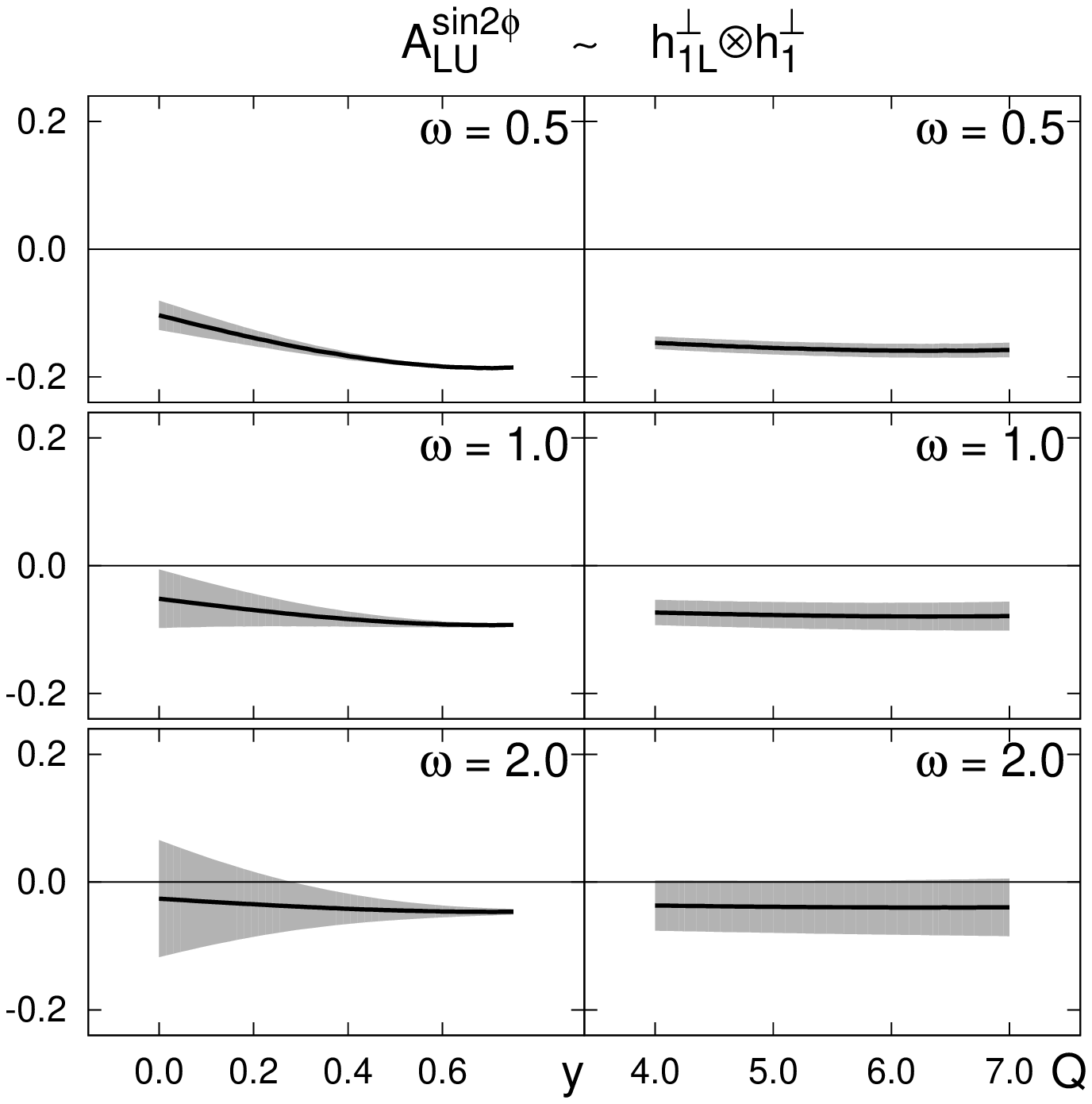}
\caption{\label{fig:e906} Azimuthal asymmetries $A_{TU}^{\sin(2\phi-\phi_S)}$  (left panels) , $A_{TU}^{\sin(2\phi+\phi_S)}$ (central panels), and $A_{LU}^{\sin2\phi}$ (right panels) at E906.}
\end{center}
\end{figure*}

\item E906

There is a new proposal to use proton beams from the main injector
at $E_p=120~\mathrm{GeV}$ to collide on the polarized proton target
(NH$_3$) by E906 Collaboration~\cite{Liu:2010kb} at Fermi Lab. The
polarized dimuon Drell-Yan program at E906 might be applied to
measure the asymmetries defined in Eqs.~(\ref{asin2m}),
(\ref{asin2p}) and (\ref{asin2}). The E906 kinematics are given as
\[\begin{split}
&\sqrt{s}=15~\mathrm{GeV},~~0.3<x_1<0.7,~~0.1<x_2<0.3,\\
&0<q_T<1~\mathrm{GeV},~~4~\mathrm{GeV} <Q<7~\mathrm{GeV},
\end{split}\]
corresponding to $0<y<0.76$.
The calculated asymmetries $A_{TU}^{\sin(2\phi-\phi_S)}$, $A_{TU}^{\sin(2\phi+\phi_S)}$, and $A_{LU}^{\sin2\phi}$ are shown in the left, central, and right panels of Fig.~\ref{fig:e906}.

\item NICA

NICA at JINR
might realize both longitudinally and transversally polarized beams of
protons at $\sqrt{s} = 12 \sim 27~\mathrm{GeV}$~\cite{Meshkov:2011zz}. The
kinematics applied in our calculation at NICA are
\[\begin{split}
&\sqrt{s} = 27~\mathrm{GeV},~~4~\mathrm{GeV}<Q<9~\mathrm{GeV},\\
&0<q_T < 1~\mathrm{GeV},~~0.1<x_1<0.8,
\end{split}\]
corresponding to $-1.1<y<1.1$. Here we choose the highest c.m. energy to avoid the overlap with other experiments. We present the asymmetries $A_{TU}^{\sin(2\phi-\phi_S)}$, $A_{TU}^{\sin(2\phi+\phi_S)}$, and $A_{LU}^{\sin2\phi}$ in the left, central, and left panels of Fig.~\ref{fig:nica}, respectively.

\end{itemize}

\begin{figure*}[t]
\begin{center}
\includegraphics[width=0.32\textwidth]{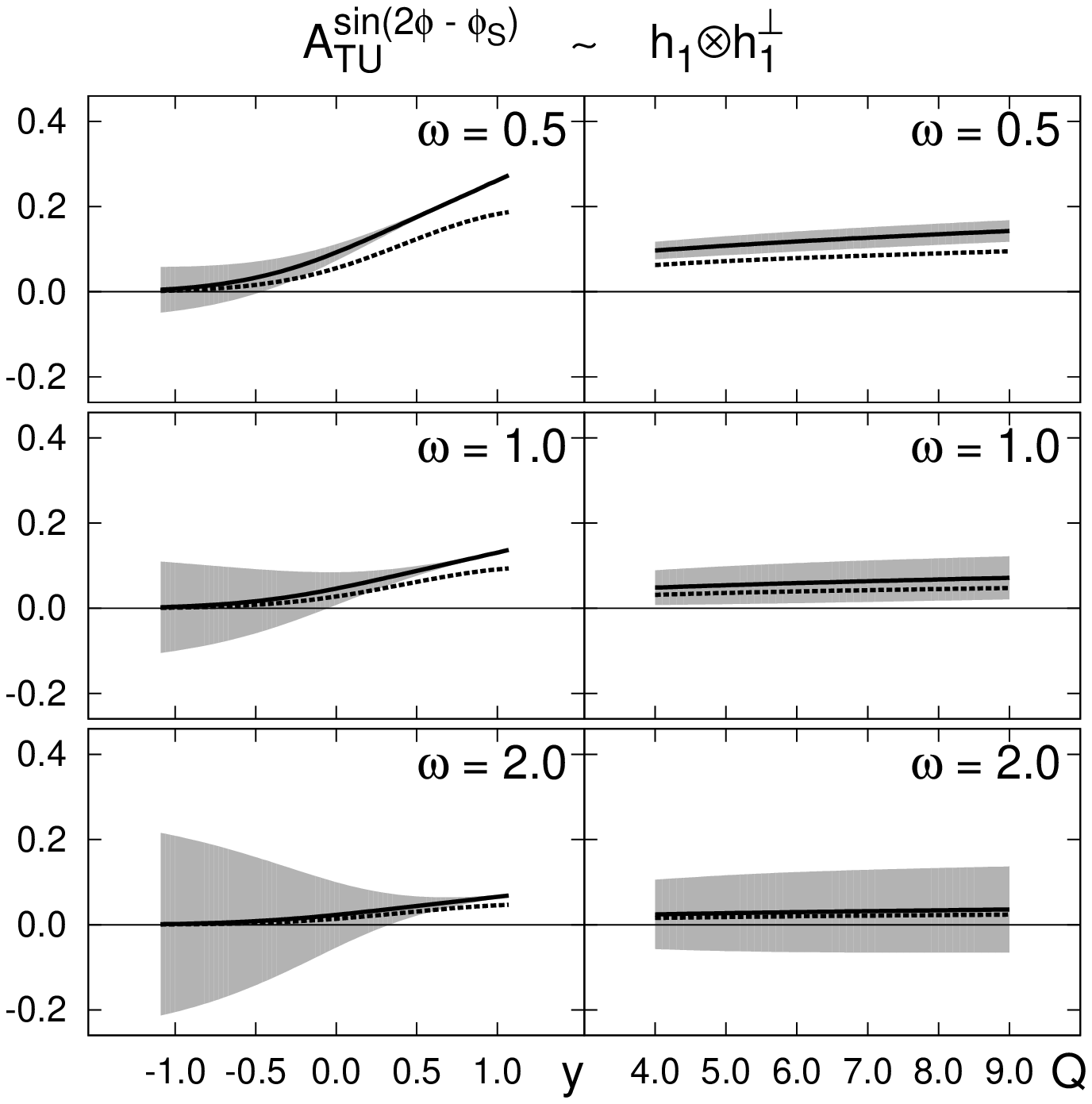}
\includegraphics[width=0.32\textwidth]{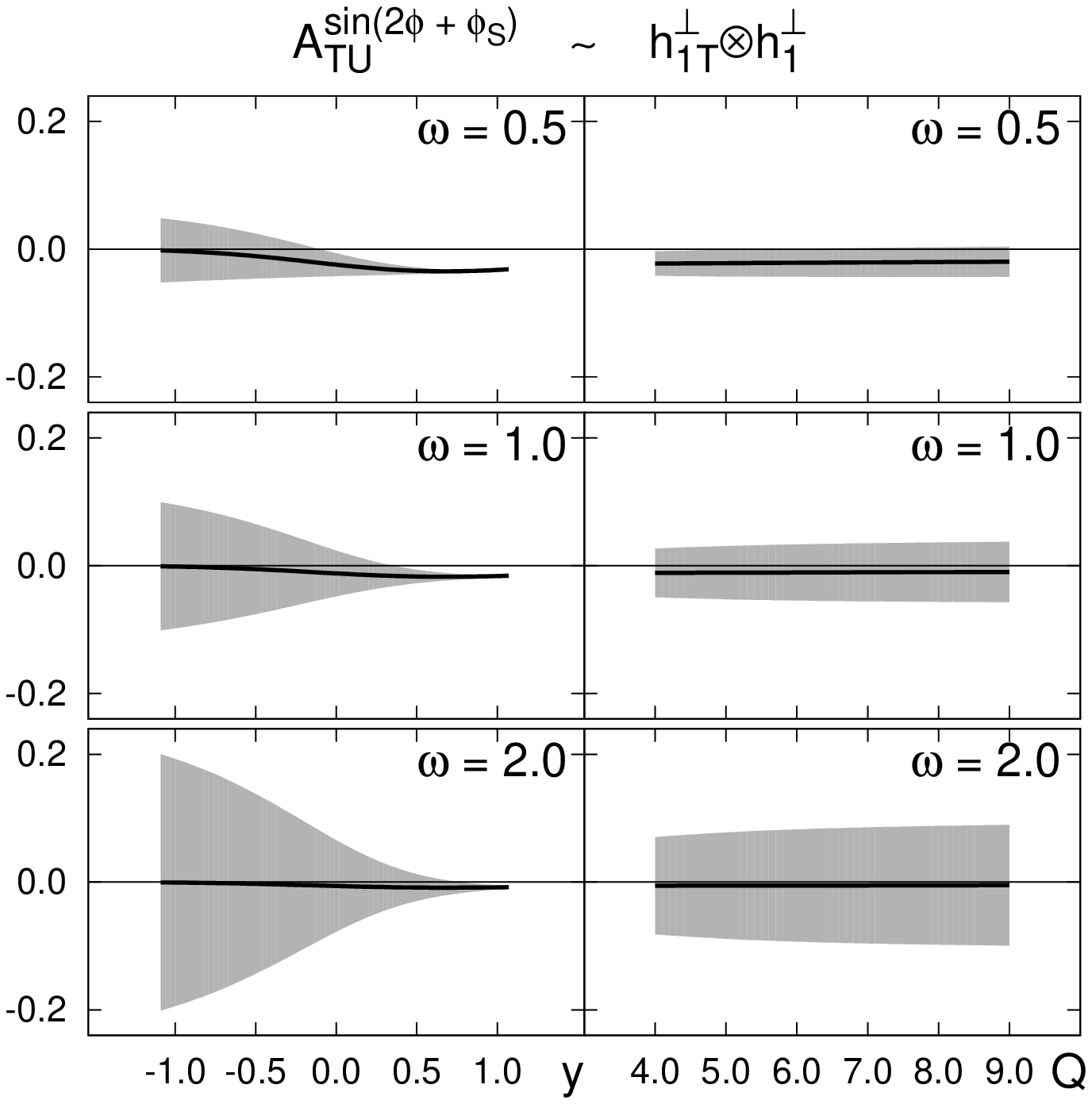}
\includegraphics[width=0.32\textwidth]{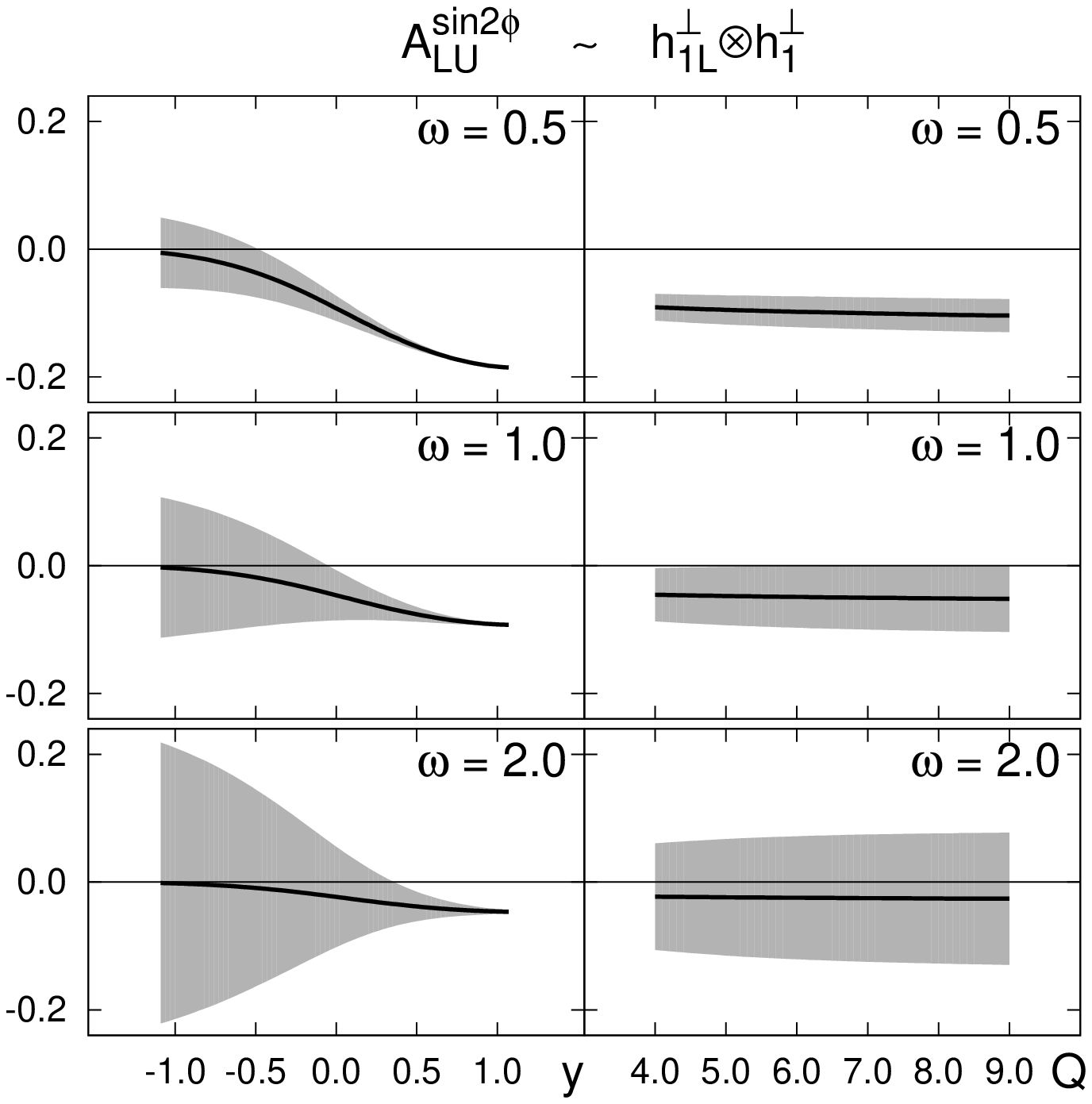}
\caption{\label{fig:nica} Azimuthal asymmetries $A_{TU}^{\sin(2\phi-\phi_S)}$ (left panels) , $A_{TU}^{\sin(2\phi+\phi_S)}$ (central panels),  and $A_{LU}^{\sin2\phi}$ (right panels) at NICA.}
\end{center}
\end{figure*}

Our predictions at RHIC, J-PARC, E906, and NICA show that the
asymmetries $A_{TU}^{\sin(2\phi-\phi_S)}$,
$A_{TU}^{\sin(2\phi+\phi_S)}$, and $A_{LU}^{\sin2\phi}$ are sensitive
to the Boer-Mulders functions of sea quarks. This can be seen by
comparing the plots in the three rows of each figure. The size of
sea content can be described by the parameter $\omega$ appearing in
the parametrizations of the Boer-Mulders functions. The case
$\omega=1$ corresponds to the central values of the Boer-Mulders
functions (which we refer as normal case), while $\omega=0.5$ corresponds to
much smaller valence and much larger sea values (large sea case),
and $\omega=2$ corresponds to much larger valence and much smaller
sea values (small sea case) compared to the central values. In the
normal case, we can see from the $Q$-dependent plots that the
asymmetries $A_{TU}^{\sin(2\phi-\phi_S)}$ and $A_{LU}^{\sin2\phi}$
are sizable at all entire allowed $Q$ regions. As $\omega$ increases
or decreases, the asymmetries decrease or increase correspondingly.
Therefore, their measurements could be used to discriminate
different scenarios of the Boer-Mulders functions.

At larger backward rapidity region (RHIC and NICA) or midrapidity
region (J-PARC and E906), the figures show that there are
uncertainties contributed by the unknown sea content of $h_1$,
$h_{1T}^\perp$, and $h_{1L}^\perp$ allowed by the positivity bounds,
especially in the small sea case. In some cases the uncertainties
are so large that the sizes and signs of the asymmetries can not
been determined. Precision measurement at these regions will provide
further constraints on the sea content of $h_1$, $h_{1T}^\perp$, and
$h_{1L}^\perp$.

The plots for  $A_{TU}^{\sin(2\phi-\phi_S)}$ and
$A_{LU}^{\sin2\phi}$ show that these asymmetries are larger at the
forward rapidity region, about $10\%$ in magnitude in the normal
case. Furthermore, our plots show that at the forward rapidity region,
the contributions from the $T$-even chiral-odd distributions
of valence quarks dominate, that is, they are less contaminated by their sea
content. Hence the asymmetries at forward rapidity are
measurable and the measurements on them are ideal to access the
valence content of $h_1$, and $h_{1L}^\perp$ at large $x$ region.

The magnitudes of the asymmetries $A_{TU}^{\sin(2\phi-\phi_S)}$ are
larger than  $A_{LU}^{\sin2\phi}$ and $A_{TU}^{\sin(2\phi+\phi_S)}$.
This is because that the size of the transversity distributions in
the light-cone quark-diquark model is larger than that of
$h_{1T}^{\perp}$ and $h_{1L}^{\perp}$. The comparison of different
types of asymmetries might be used to distinguish the sizes of different
$T$-even chiral-odd distributions and to check the approximate
relations among TMDs~\cite{Avakian:2007mv}.

\section{Conclusion}

We have studied the azimuthal asymmetries in the single polarized
proton-proton Drell-Yan processes by considering particularly the
contributions of the leading-twist chiral-odd quark distributions,
i.e., the Boer-Mulders function, transversity, pretzlosity and
longitudinal transversity. We define the azimuthal asymmetries
$A_{TU}^{\sin(2\phi+\phi_S)}$ and $A_{TU}^{\sin(2\phi-\phi_S)}$ in
transverse single polarized $p^\uparrow p$ Drell-Yan processes, and
$A_{LU}^{\sin2\phi}$ asymmetry in the longitudinal single polarized
$p^\rightarrow p$ Drell-Yan processes. Using the predictions for the
transversity, pretzlosity and longitudinal transversity from the
light-cone quark-diquark model, and the Boer-Mulders functions
extracted from the unpolarized Drell-Yan data at low transverse
momentum, we present a comprehensive phenomenological analysis of
the asymmetries $A_{TU}^{\sin(2\phi+\phi_S)}$,
$A_{TU}^{\sin(2\phi-\phi_S)}$, and $A_{LU}^{\sin2\phi}$  at RHIC,
J-PARC, E906, and NICA. In all these facilities there are polarized
Drell-Yan programs in preparation or being planned, including
collider experiments (RHIC and NICA) and fixed-target experiments
(RHIC, J-PARC, and E906). Our study shows that the polarized
Drell-Yan programs at various facilities can be used to explore
the valence and sea content of the leading-twist chiral-odd
distributions in wide kinematical regions.

\section*{Acknowledgement}
This work is partially supported by National Natural Science
Foundation of China (Grants No.~10905059, No.~11005018,
No.~11021092, No.~10975003, No.~11035003, and No.~11120101004) and
by FONDECYT (Chile) under Project No.~11090085.

\end{document}